\documentclass[12pt,twocolumn,preprint,tighten,trackchanges]{aastex63}
\usepackage{natbib}
\usepackage{multirow,color}
\usepackage{bbding}
\usepackage{amssymb}
\usepackage{ulem}
\usepackage{color,xcolor}
\usepackage{longtable}
\usepackage{array}
\usepackage{bm}
\usepackage{url}
\usepackage{graphicx}

\def\gsim{\;\lower4pt\hbox{${\buildrel\displaystyle >\over\sim}$}\;}
\def\lsim{\;\lower4pt\hbox{${\buildrel\displaystyle <\over\sim}$}\;}
\def\grls{\;\lower4pt\hbox{${\buildrel\displaystyle >\over <}$}\;}

\definecolor{darkcyan}{RGB}{0,102,204}

\bibpunct{(}{)}{,}{o}{}{,}

\begin{document}
\title{The Source Locations of Major Flares and CMEs in the Emerging Active Regions} 

\author{Lijuan Liu}
\affiliation{Planetary Environmental and Astrobiological Research Laboratory (PEARL), School of Atmospheric Sciences, Sun Yat-sen University, Zhuhai, Guangdong, 519082, China}
\affiliation{CAS Key Laboratory of Geospace Environment, Department of Geophysics and Planetary Sciences, University of Science and Technology of China, Hefei, Anhui, 230026, China}
\affiliation{CAS center for Excellence in Comparative Planetology, China}
{\correspondingauthor{Lijuan Liu}}
{\email{liulj8@mail.sysu.edu.cn}}

\author{Yuming Wang}
\affiliation{CAS Key Laboratory of Geospace Environment, Department of Geophysics and Planetary Sciences, University of Science and Technology of China, Hefei, Anhui, 230026, China}
\affiliation{CAS center for Excellence in Comparative Planetology, China}

\author{Zhenjun Zhou}
\affiliation{Planetary Environmental and Astrobiological Research Laboratory (PEARL), School of Atmospheric Sciences, Sun Yat-sen University, Zhuhai, Guangdong, 519082, China}
\affiliation{CAS center for Excellence in Comparative Planetology, China}

\author{Jun Cui}
\affiliation{Planetary Environmental and Astrobiological Research Laboratory (PEARL), School of Atmospheric Sciences, Sun Yat-sen University, Zhuhai, Guangdong, 519082, China}
\affiliation{CAS Key Laboratory of Lunar and Deep Space Exploration, National Astronomical Observatories, Chinese Academy of Sciences, Beijing, 100012, China}

\begin{abstract}

Major flares and coronal mass ejections (CMEs) tend to originate from the compact polarity inversion lines (PILs) in the solar active regions (ARs). 
Recently, a scenario named as ``collisional shearing'' is proposed by \citet{Chintzoglou_2019} to explain the phenomenon, 
which suggests that the collision between different emerging bipoles is able to form the compact PIL, driving the shearing and flux cancellation that are responsible to the subsequent large activities. 
In this work, through tracking the evolution of 19 emerging ARs from their birth until they produce the first major flares or CMEs, 
we investigated the source PILs of the activities, i.e., the active PILs,  
to explore the generality of ``collisional shearing".  
We find that none of the active PILs is the self PIL (sPIL) of a single bipole. 
We further find that 11 eruptions originate from the collisional PILs (cPILs) formed due to the collision between different bipoles, 6 from the conjoined systems of sPIL and cPIL,  
and 2 from the conjoined systems of sPIL and ePIL (external PIL between the AR and the nearby preexisting polarities). 
Collision accompanied by shearing and flux cancellation is found developing at all PILs prior to the eruptions, with $84\%$ (16/19) 
cases having collisional length longer than 18~Mm. 
Moreover, we find that the magnitude of the flares is positively correlated with the collisional length of the active PILs,  indicating that the intenser activities tend to originate from the PILs with severer collision. 
The results suggest that the ``collisional shearing'', 
i.e., bipole-bipole interaction during the flux emergence is a common process in driving the major activities in emerging ARs.

\end{abstract}

\section{INTRODUCTION}\label{sec:intro}

Solar flares and coronal mass ejections (CMEs) are among the most violent activities in the solar atmosphere. 
Their main producers are known to be the solar active regions (ARs). 
It is generally accepted that the ARs are formed by the magnetic flux emerging from the solar interior. 
The emergence of a single $\Omega$-shaped flux tube can form the simplest bipolar region, 
the two polarities of which are the intersections between the tube axial field and the photosphere~\citep[][and reference therein]{Schmieder_2014}. 
During the emergence, the two main polarities move apart, 
with typical signatures such as small moving dipoles (small opposite-sign polarity pairs) appearing between them~\citep[e.g.,][]{Strous_1999, Bernasconi_2002, Centeno_2012}.  
Different bipoles can interact to form complex configuration, 
such as quadrupolar configuration.

Observationally, not all ARs are able to generate large flares or CMEs. 
It is found that the eruption-producing ARs tend to be larger, 
containing larger amount of magnetic free energy, hosting more complex configuration than a single bipole~\citep[e.g.,][]{Falconer_2002, Leka_Barnes_2003a, Falconer_2006, Leka_Barnes_2007, Georgoulis_Rust_2007, Falconer_2008, Chen_Wang_2011, Lliu_2016}. 
Enough free energy is only a necessary condition for the AR to produce major flares or CMEs. 
The trigger of the activities may involve more complex process such as magnetohydrodynamics (MHD) instabilities~\citep[e.g.,][]{Torok_2004, Kliem_2006} and magnetic reconnection~\citep[e.g.,][]{Antiochos_1999, Moore_2001}. 
Considering the evolution of the ARs, 
their ability to produce eruptions increases as the flux emergence proceeds~\citep[][and reference therein]{Van_2015}. 
Although the decaying ARs can also produce eruptions mainly due to flux cancellation (magnetic reconnection near the photosphere), 
the most violent activities tend to originate from the still emerging and evolving ARs~\citep{Schrijver_2009}.

The lines where the polarities change signs are called polarity inversion lines (PILs). 
It is found that within the eruption-producing ARs, 
compact PILs, i.e., the PILs with high spatial gradient, 
are usually the sources of major flares and CMEs~\citep{schrijver_2007_characteristic}. 
Shearing motions and sunspot rotations can always be found near the compact PILs. 
Those motions are suggested to be able to shear or twist the field lines so that to inject the free energy and magnetic helicity into the AR~\citep[e.g.,][]{Fanyh_2009, Yanxl_2015}, 
being closely related to the eruptions. 
Converging motions and flux cancellation are also frequently observed~\citep[e.g.,][]{Green_2011, Cheng_2014a}. 
In the classical flux cancellation model of the single bipolar region~\citep{VanBallegooijen_1989}, 
converging motions can bring the opposite-sign footpoints of different loops of the sheared arcade together, 
leading to flux cancellation which forms the flux rope (a set of helical field lines winding around a common axis). 
The subsequent eruption of the flux rope may form the CME, 
and generate the flare through magnetic reconnection
~\citep[e.g.,][]{Shibata_1995}.

Combining the shearing motion, converging motion and flux cancellation in the emerging ARs, 
\citet{Chintzoglou_2019} proposed a new scenario to explain the compact PIL formation and the origin of the major solar activities. 
Through tracking the evolution of two flare- and CME-productive ARs from the very beginning of their emergence, the authors found that the clusters of the eruptions 
correlated well with the onset of a process named as ``collisional shearing''. 
During the ``collisional shearing'', different bipoles appear on the photosphere due to the emergence of different flux tubes. 
For each bipole, the two main polarities move apart as the two legs of the flux tube separate during the emergence. The separation results in converging motions, thus collision, between the nonconjugated opposite-sign polarities, 
forming the compact PIL. 
The continuous collision further drives shearing and flux cancellation, leading to the magnetic reconnection and the formation of flux ropes, followed by a series of large activities. 
This kind of compact PIL formed due to collision between non-conjugated polarities is defined as collisional PIL (cPIL), differentiating from the self PIL (sPIL) formed between the conjugated polarities~\citep[all defined in][]{Chintzoglou_2019}. 
The authors 
pointed out that the overall PIL in the AR was naturally an integral system of sPIL and cPIL. 
They further identified two types of collisional shearing patterns, case A and case B. In case A, the two bipoles emerge simultaneously, driving the collision by self-separation. In case B, the two bipoles emerge sequentially, followed by the collision.  
On the contrast to the eruption models considering the bipolar ARs~\citep[e.g.,][]{VanBallegooijen_1989, Fanyh_2001}, 
the bipole-bipole interaction plays the major role in the collisional shearing scenario. 

\citet{LLiu_2019} also reported that during the early emergence phase of NOAA AR 12673, a magnetic flux rope above the AR's central PIL was formed through flux cancellation and shearing during the ``collisional shearing''. 
They further suggested that the subsequent recurrent eruptions and flux ropes reformation were driven by the same process. 
Both works indicate that the ``collisional shearing'' may be important in driving large solar activities in emerging ARs.  
However,  
no statistical study on this phenomenon has been done according to our best knowledge. 
In this work,  
we perform a statistical research on the evolution of 19 ARs from their birth until they produce the first major activity, i.e., a large flare ($\geq$M1.0 class) or a CME, 
to explore the generality of the ``collisional shearing''. 
We mainly focus on the properties of the source PILs of those activities, 
which are named as active PILs in the following.  

The paper is organized as follows. In Section~\ref{sec:obs_ana}, we describe the event sample, the data and the method we use. 
The statistical result and typical examples are presented in Section~\ref{sec:result}.  
We give the discussions and conclusions in Section~\ref{sec:discuss}.


\section{Observation and DATA Analysis}\label{sec:obs_ana}

The 19 ARs we studied are selected from the list presented in~\citet{Kutsenko_2019}, 
which contains 423 emerging ARs observed by the Helioseismic and Magnetic Imager~\citep[HMI;][]{Scherrer_2012} on board {\it Solar Dynamics Observatory} ({\it SDO;}~\citealt{Pesnell_2012}). 
Our ARs fulfill the following criteria: 
the AR should emerge and generate at least one major flare ($\geq$M1.0 class, either confined or not) or CME within the region between Stonyhurst longitudes 60$^{\circ}$E and 60$^{\circ}$W (roughly having disk-centered angle $\Theta \leq 60^{\circ}$).   
The criteria ensure a relatively high signal-to-noise ratio of the HMI data, as the data taken near the solar limb are subjected to severe uncertainties and projection effect. 
We check the flares in the soft X-ray (SXR) flare catalog of the {\it Geostationary Operational Environmental Satellite (GOES)}\footnote{\url{https://www.ngdc.noaa.gov/stp/space-weather/solar-data/solar-features/solar-flares/x-rays/goes/xrs/}\label{ft:f1}}. 
The CMEs are examined in the 
{\it Solar and Heliospheric Observatory (SOHO)}/Large Angle and Spectrometric Coronagraph (LASCO) CME catalog maintained at the Coordinated Data Analysis Workshop (CDAW) data center\footnote{\url{https://cdaw.gsfc.nasa.gov/CME_list/}}~\citep{Gopalswamy_2009}, 
and the {\it Solar Terrestrial Relations Observatory (STEREO)}/Sun–Earth Connection Coronal and Heliospheric Investigation (SECCHI) COR2 CME catalog that is recorded by the Solar Eruptive Event Detection System (SEEDS) maintained by the George Mason University\footnote{\url{http://spaceweather.gmu.edu/seeds/secchi.php}}~\citep{Olmedo_zhang_etal_2008}. 
The basic properties of the ARs and their first major activities are given in Table~\ref{tb:props}.

We use a data product from  the {\it SDO}/HMI called Spaceweather HMI Active Region Patches~\citep[SHARPs;][]{Hoeksema_etal_2014, Bobra_2014} to track the evolution of the ARs. The SHARP data series produces cutout maps 
of the automatically tracked ARs for their entire transit, 
providing the photospheric vector magnetic field as well as the line-of-sight (LOS) magnetograms with a spatial resolution of 0\farcs 5 and a time cadence of 12 minutes. 
A specific version of the SHARP data, which is remapped from the CCD coordinates to the heliographic Cylindrical Equal-Area (CEA) projection coordinates, is used. 
To locate the exact source PILs of the first major activities, 
we inspect the extreme ultraviolet (EUV) and ultraviolet (UV) images provided by the Atmospheric Imaging Assembly~\citep[AIA;][]{Lemen_etal_2012} on board {\it SDO}. 
The images have a spatial resolution of 0\farcs 6 and a time cadence up to 12 seconds. 
We mainly use the hot passband 131 \AA~($\sim$10~MK) and the cool passband 304 \AA~($\sim$50000K) to show the eruption details. 
If the eruption signatures are not clear enough in the two passbands, the 211 \AA~passband ($\sim$2.0 MK) is further used to show the possible coronal dimmings associated with the CMEs. 

We follow the methods employed in ~\citep{Chintzoglou_2019} to quantify the ``collisional shearing'' process. 
Specifically, we calculate the magnetic flux of the AR, track all polarities involved, and identify the collisional portions of the active PILs using the SHARP data. 
The radial component of the vector field, $B_r$, is a natural choice to do the analysis. 
However, the noise level of the vector field is as high as 100~G~\citep{Hoeksema_etal_2014}, 
and is severe in the regions beyond $30^{\circ}$ from the disk center. 
The other choice is to use the low noise (as low as 10~G) 
radialized LOS magnetic field data, 
i.e, the LOS data corrected by dividing the cosine of the angle between LOS and local normal of the disk 
($\displaystyle B_{LOS}=B_{LOS}^{raw}/\cos\theta$, in which $B_{LOS}^{raw}$ is the raw LOS data). 
In the following, the mentioned $B_{los}$ 
indicates the data after correction.  
Nevertheless, the correction assumes that 
the horizontal component of the magnetic field ($B_h$) contributes much less to $B_{los}$ than 
the vertical component ($B_r$), 
which may not hold when the AR is not near the disk center. 
The uncertainty introduced by the correction is hard to be estimated without low noise $B_h$.   
Overall, the two datasets ($B_r$ and $B_{los}$) are safe to be used in the region having disk-centered angle $\leq 30^{\circ}$, 
but have different flaws out of the region. 
We thus repeat the analysis on both datasets for comparison.

The flux-weighted centroid for each polarity is calculated within a radius of 5~Mm (about 14 pixels) around its peak intensity~\citep[used in][]{Chintzoglou_2019}.  
Note that, some like-signed polarities from different bipoles may be extremely close to each other, having inseparable boundaries, 
thus are called as a group~\citep[defined in][]{Chintzoglou_2019}. Based on the time-series centroids, we plot the trajectory of each centroid and the evolution of distances between different centroids.  

We further quantify the collision strength and the non-potential shear of the active PILs. 
The collision strength is characterized by the length of the collisional, i.e., compact, high spatial gradient part of the PIL 
that is obtained by the method described in~\citet{schrijver_2007_characteristic} and \citet{Chintzoglou_2019}.   
For the polarities forming the active PIL, the strong field regions (kernels) are firstly isolated with various thresholds,  
then dilated by the kernel with a size of 5 pixels~\citep[the size used in][]{Chintzoglou_2019}.  
The region where the positive and negative polarities intersect is taken as the primary collisional PIL part, same as in~\citet{Chintzoglou_2019}.    
A thinning operation is further performed 
until the part is one-pixel wide~\citep[see details in][]{Chintzoglou_2019}, 
which is taken as the final collisional PIL part. 
All pixels of the collisional part are summed up to get its length, $L_{cPIL}$. 
Here various thresholds are used to avoid the bias, 
which are 100~G, 125~G, 150~G and 175~G for $B_r$, and are 50~G, 75~G, 100~G, 125~G, 150~G and 175~G for $B_{los}$. 
As the collision is detected through the dilation of strong field kernel, 
the threshold lower than 50~Gauss may not fit the definition of strong field, 
and larger than 175~Gauss may unnecessarily reduce the collisional length. 
The reason for not using 50~G and 75~G for $B_r$ is that they are lower than the noise level of $B_r$. 
It is found that all thresholds yield the similar results (see Section~\ref{subsec:stat}). We thus present the results of $B_r$ at 100~Gauss~\citep[threshold used in][]{Chintzoglou_2019} for simplicity when showing the examples.

The non-potential shear is characterized by the mean shear angle $S$ of the field at the collisional PIL part. 
To calculate the mean shear angle, we firstly compute the potential field with the frourier transformation method~\citep{Alissandrakis_1981} using photospheric $B_r$ 
as the input, 
then calculate the mean shear angle $S$ by the equation $\displaystyle S=\frac{1}{N}\Sigma arccos(\frac{\mathbf{B}^{Obs}\cdot \mathbf{B}^{pot}}{|B^{Obs}||B^{Pot}|})$~\citep{Bobra_2014}, 
in which $N$ is the number of the pixels of the collisional part of the PIL, 
$\mathbf{B}^{Obs}$ is the observed vector magnetic field, and $\mathbf{B}^{pot}$ is the extrapolated potential field. 
For the LOS field dataset, $S$ is calculated in the same way using the collisional pixels identified on $B_{los}$ maps. 

We further inspect the difference between the parameters obtained on $B_r$ and $B_{los}$ datasets. 
For two values calculated on different datasets at the same time, their percentage difference 
is measured as $\displaystyle r=\frac{p_{los}-p_r}{p_r}\times 100\%$, 
in which $p_{los}$ and $p_r$ indicate $L_{cPIL}$ or $S$ obtained on $B_{los}$ and $B_r$, respective. 
The overall difference is then calculated as the mean value of the percentage differences of all data by $\displaystyle \frac{1}{N}{\Sigma r}$, 
in which $N$ is the number of data points from the flux emergence start to the flare onset.

\section{Results}\label{sec:result}

We track the ARs and locate the active PILs where the first major activities originate. 
For an AR, 
except the sPIL and cPIL~\citep[defined in][]{Chintzoglou_2019}, there may be another type of PIL 
formed between the AR and the nearby preexisting polarities, 
defined as the external PIL \citep[ePIL,][]{MacKay_2008, Chintzoglou_2019}. 
We find that the active PILs in our sample could be classified into three types: 
cPIL, 
of which the PILs are complete collisional PILs; 
the conjoined sPIL/cPIL, 
in which the PILs are the combination of the self PILs and collisional PILs; 
the conjoined sPIL/ePIL, 
in which the PILs are the integral 
systems of the self PILs and external PILs.  
Finally, we get 11 cPILs, 
6 conjoined sPIL/cPIL 
and 2 conjoined sPIL/ePIL. 
Their information is shown in Table~\ref{tb:props}. 
Eight typical examples are given in the following.

\subsection{Examples of cPILs}\label{subsec:cpil} 

\subsubsection*{The cPIL in NOAA AR 11162}\label{susubbsec:cpil_11162}

\begin{figure*}
\begin{center}
\epsscale{1.1}
\plotone{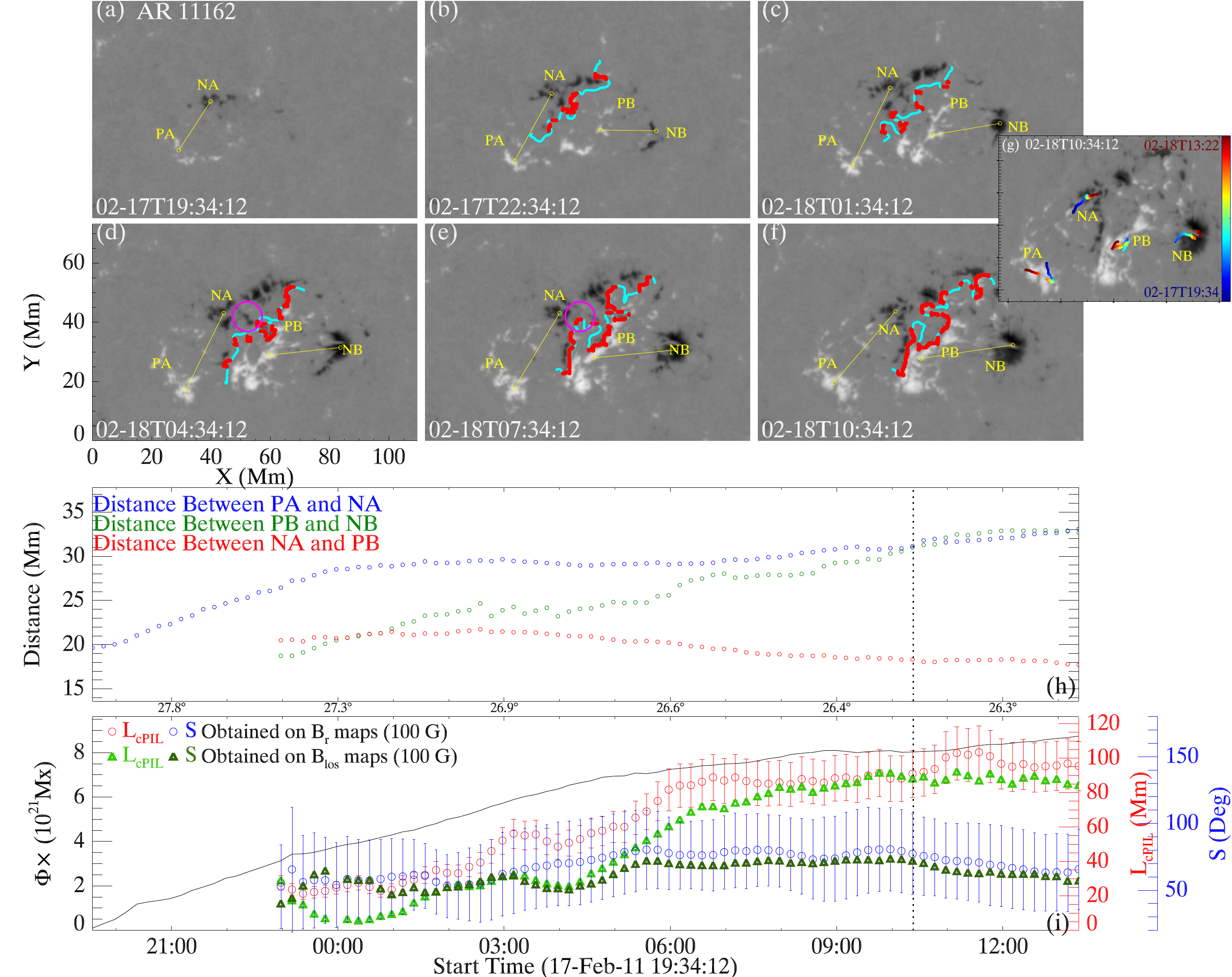}
\caption{
Evolution of NOAA AR 11162. 
(a)-(f) Evolution of the photospheric $B_r$.  
White (black) patches are the positive (negative) polarities, saturating at $\pm 2000$ Gauss. PA and NA indicate the positive and negative polarities of bipole A, while PB and NB are for bipole B. The cyan lines mark the source PIL of the first major activity, which are obtained by the direct contours at $B_r=0 $. The red lines indicate the collisional components of the PILs.  The collisional parts are obtained by the method described in Section~\ref{sec:obs_ana}. 
The yellow circles mark the flux-weighted centroids of the polarities, with those 
of the conjugated polarities connected by the yellow lines. 
The magenta circles in panels (d) and (e) mark the location where a patch of negative polarity disappears. 
An animation of the magnetograms lasting from 2011-02-17T14:10 to 2011-02-18T12:58 is available online. (g) The trajectories of the flux-weighted centroids of all polarities tracked on $B_r$ dataset. As time elapses, the color of the dots changes from blue to red. (h) Evolution of the distances between each pair of conjugated polarities and between the colliding, nonconjugated polarities. (i) Evolution of the magnetic flux ($\Phi$), the length ($L_{cPIL}$) and mean shear angle ($S$) of the collisional part of the PIL obtained on both $B_r$ and radialized $B_{los}$ datasets at the threshold of 100~G. Errors for the $L_{cPIL}$ from $B_r$ are further shown, accounting for 15$\%$ 
of the values of $L_{cPIL}$. The percentage is estimated by \citet{Chintzoglou_2019} using the same cPIL detection method. The errors for $S$ are the standard deviations of the shear angles of the collisional parts of the PILs. Vertical lines in (h) and (i) mark the onset instant of the flare. 
The 
locations at the top of panel (i) are disk-centered angles of the AR 
at the relevant timings.  
}\label{fig:11162_b}
\end{center}
\end{figure*}

NOAA AR 11162 is composed of two bipoles emerging sequentially. Its collisional shearing case is case B. 
From the the flux emergence start to the flare onset, the AR transits from Stonyhurst longitude 10$^\circ$E to 0$^\circ$E, with its disk centered angle changing from $28.1^{\circ}$ to $26.3^{\circ}$. 
The evolution of the AR is shown in Figure~\ref{fig:11162_b},  
in which the $B_r$ magnetograms, the polarities centroids and their distances obtained on $B_r$, and the length and shear angle of the collisional PIL parts detected at 100~G on both $B_r$ and $B_{los}$ datasets are displayed. 
We determine the conjugated polarities of a bipole according to three criteria: 
firstly, the two polarities emerge at the same time; 
secondly, they move apart from each other; 
thirdly, typical signatures such as moving dipoles appear between them.

The first bipole, named as bipole A, starts to emerge at around 2011-02-17T15:34. Four hours later, the other bipole, bipole B emerges at the west of bipole A, with its positive polarity (PB in Figure~\ref{fig:11162_b}) locating close 
to the negative polarity of bipole A (NA in Figure~\ref{fig:11162_b}). As emergence goes on, the conjugated polarities of each bipole separate, rendering the nonconjugated NA and PB to approach each other. 
The PIL between NA and PB is the active PIL (cyan line in Figure~\ref{fig:11162_b}). 
On the PIL, 
collisional signature appears and grows gradually (red line part in Figure~\ref{fig:11162_b}). Clear disappearance of the negative polarities is observed near the PIL (enclosed in magenta circle in Figure~\ref{fig:11162_b}), indicating the flux cancellation. We further track the motion of the flux-weighted centroids of all polarities and show their trajectories in Figure~\ref{fig:11162_b}(g). The centroid of NA moves northwestward, while that of PA firstly moves southwestward then northeastward. 
Meanwhile, the centroid of NB moves westward, while that of PB moves eastward. NA and PB slightly approach and shear against each other. 
Since the active PIL is formed by the collision of nonconjugated polarities, we classify it as a cPIL. 

The evolution of the distances between the polarities further proves the ``separation and collision'' process (Figure~\ref{fig:11162_b}(h)). 
From the flux emergence start to the flare onset, both the distances between the conjugated polarities of biople A and bipole B increase. The former increases from 19.8~Mm to 31.1~Mm, while the latter grows from 18.7~Mm to 31.0~Mm. 
The distance between NA and PB slightly decreases from 20.5~Mm to 18.3~Mm. 
When the first major activity occurs, 
the AR grows into a medium sized region with the unsigned magnetic flux of $8.0\times 10^{21}$~Mx (Figure~\ref{fig:11162_b}(i)). 
The length of the collisional PIL part ($L_{cPIL}$) obtained on $B_r$ series has increased from 27.8~Mm to 91.0~Mm, 
whilst the mean shear angle ($S$) increases from 64.1$^{\circ}$ to 77.3$^{\circ}$. 
The $L_{cPIL}$ and $S$ obtained on $B_{los}$ series show the similar trend. 
The former increase from 18.5~Mm to 88.3~Mm, and the latter increases from 54.4$^{\circ}$ to 72.2$^{\circ}$. 
The overall difference between $L_{cPIL}$ on the two datasets is $-24.5\%$, and that for $S$ is $-10.0\%$, 
indicating both parameters obtained on $B_{los}$ are overall smaller than those obtained on $B_r$ at the same threshold in this event. 

\begin{figure*}
\begin{center}
\epsscale{1.25}
\plotone{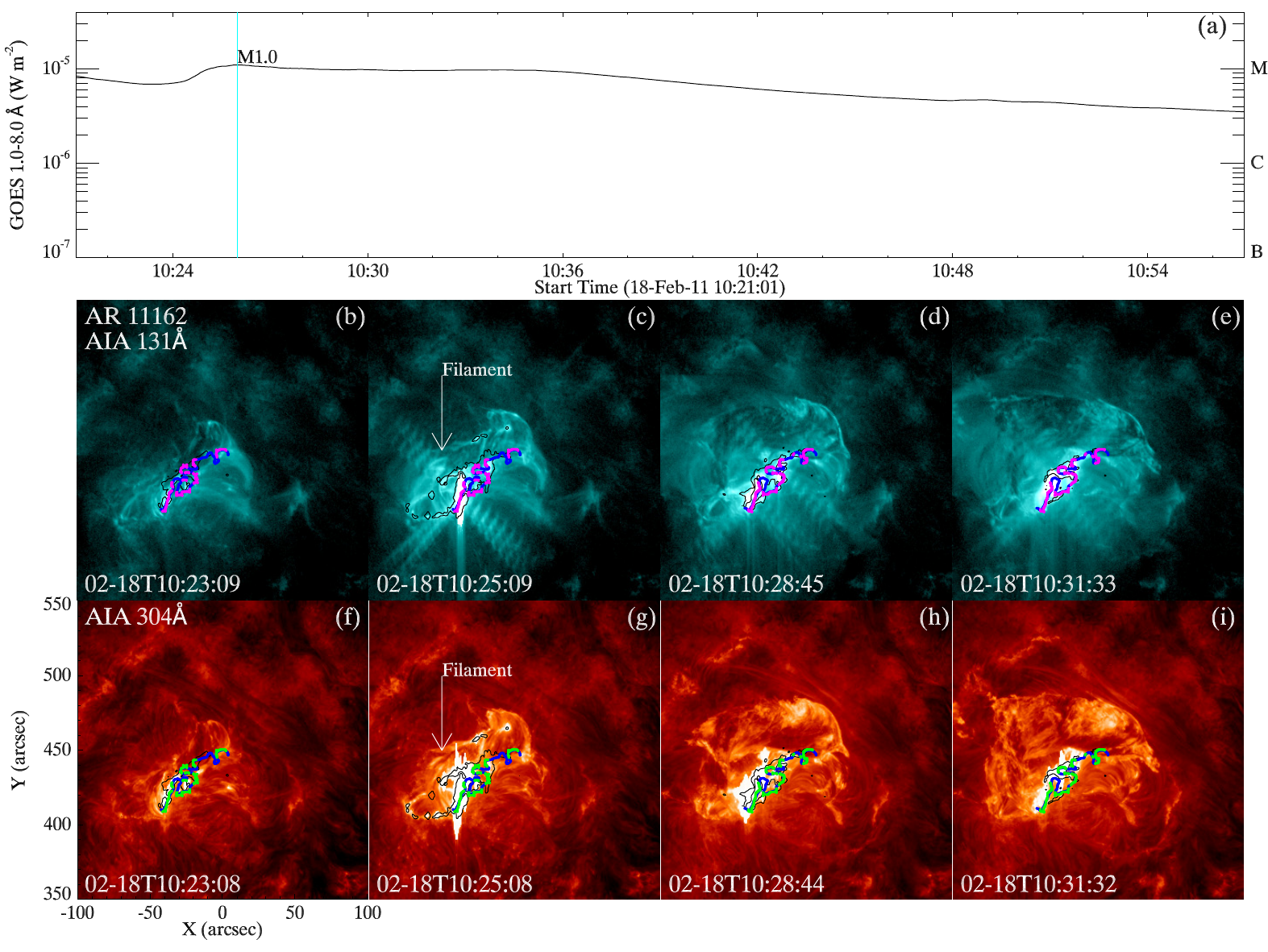}
\caption{ 
The first major activity occurred in NOAA AR 11162. (a) GOES 1-8~\AA~flux. The vertical line indicates the instant of the peak of the flare. 
(b)-(e) Eruption details captured by the AIA 131~\AA~passband. (f)-(i) Eruption details observed by the AIA 304~\AA~passband. 
Blue lines in (b)-(i) indicate the source PIL of the eruption, i.e., the PIL between NA and PB as showed in Figure~\ref{fig:11162_b}. Purple lines in (b)-(e) and green lines in (f)-(i) mark the collisional part of the PIL. The black contours in panels (b)-(i) outline the flare ribbons in the 1600~\AA~passband. 
An associated animation lasting from 2011-02-18T10:23 to 2011-02-18T10:46 is available online.  
}\label{fig:11162_flr}
\end{center}
\end{figure*}

The first major eruption from the AR is an M1.0 class flare accompanied by the failed eruption of a filament which launches at 2011-02-18T10:23 (Figure~\ref{fig:11162_flr}). 
In both the hot and cool AIA channels, 131~\AA~and 304~\AA~passbands, the eruption of the filament is observed occurring above the cPIL between NA and PB (blue lines in Figure~\ref{fig:11162_flr}). The brightenings at the PIL, and the flaring ribbons occurring at both sides of the cPIL in the 1600~\AA~passband (black contours in Figure~\ref{fig:11162_flr}) confirm that the cPIL is the source of the eruption.

\subsubsection*{The cPIL in NOAA AR 11891}\label{susubbsec:cpil_11891}

\begin{figure*}
\begin{center}
\epsscale{1.25}
\plotone{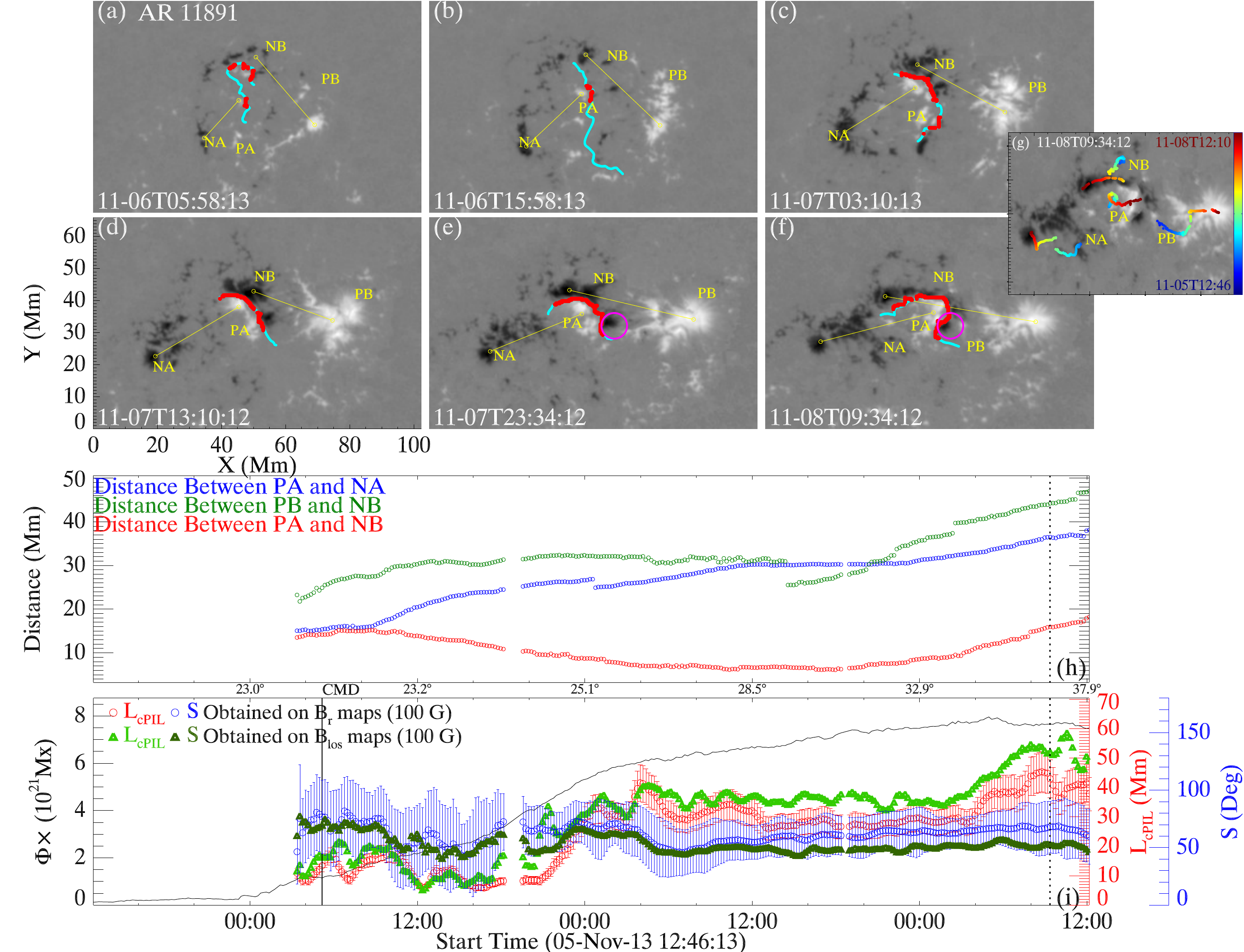}
\caption{
The evolution of NOAA AR 11891. Same layout as Figure~\ref{fig:11162_b}. An animation of the magnetograms lasting from 2013-11-05T12:58 to 2013-11-08T12:10 is available online. 
The solid vertical line labeled as ``CMD'' marks the instant when the AR passes the central meridian.}\label{fig:11891_b}
\end{center}
\end{figure*}

NOAA AR 11891 starts to emerge from around 2013-11-05T12:48. 
It transits from the Stonyhurst longitude 09$^{\circ}$E to 31$^{\circ}$W until the first major activity, with the disk centered angle changing from $24.5^{\circ}$ to $36.8^{\circ}$. 
The evolution of the AR is shown in Figure~\ref{fig:11891_b}. 
Two bipoles, named as bipole A and bipole B, appear on the photosphere simultaneously. Its collisional shearing case belongs to case A. 
The conjugated polarities of each bipole move apart from each other, 
while the positive polarity of bipole A (PA) and the negative polarity of bipole B (NB) approach each other (see Figure~\ref{fig:11891_b}(a)-(f) and the associated movie). 
The PIL between the two nonconjugated polarities is the active PIL 
(cyan lines in Figure~\ref{fig:11891_b}(a)-(f)). 
Collisional signatures gradually appear on the PIL (red line parts in Figure~\ref{fig:11891_b}(a)-(f)). 
It is thus classified as a cPIL. 
During the collision, the conjugated polarities of the two bipoles keep separating, so that the nonconjugated polarities PA and NB shear against each other. Shrinkage of the negative polarities is observed near the PIL (enclosed in magenta circles in Figure~\ref{fig:11891_b}), indicating the flux cancellation.  
The trajectories of the flux-weighted centroids of the polarities confirm the above process 
(Figure~\ref{fig:11891_b}(g)).  
PA moves northwestward while NA moves northeastward. PB moves westward while NB moves southeastward. The nonconjugated PA and NB firstly approach and then slide away from each other. The slight discontinuity of the trajectories is resulted from the jump change of the positions of the centroids. 

Until the flare, the distance between PA and NA increases from 15.1~Mm to 36.5~Mm, 
and that between PB and NB increases from 23.3~Mm to 44.2~Mm, 
confirming the separation of the conjugated polarities (Figure~\ref{fig:11891_b}(h)). 
The distance between PA and NB firstly decreases from 13.5~Mm to 6.1~Mm, 
then increases to 16.0~Mm, 
consistent with their collision and shearing away motion. 
The unsigned magnetic flux of the AR increases to $7.7\times 10^{21}$~Mx (Figure~\ref{fig:11891_b}(i)). 
The length of the collisional PIL part obtained on $B_r$ series increases from 10.9~Mm to 43.0~Mm and the mean shear angle increases from 60.2$^{\circ}$ to 68.1$^{\circ}$. The $L_{cPIL}$ obtained on $B_{los}$ also increases from 12.8~Mm to  52.0~Mm, while $S$ slightly decreases from 60.0$^{\circ}$ to 52.8$^{\circ}$. 
The overall difference of $L_{cPIL}$ from the two datasets is $25.3\%$, while that of $S$ is $-10.9\%$, 
indicating that in this case, compared to the values obtained from $B_r$, the $L_{cPIL}$ obtained from $B_{los}$ is larger 
whilst $S$ is smaller. 

\begin{figure*}
\begin{center}
\epsscale{1.25}
\plotone{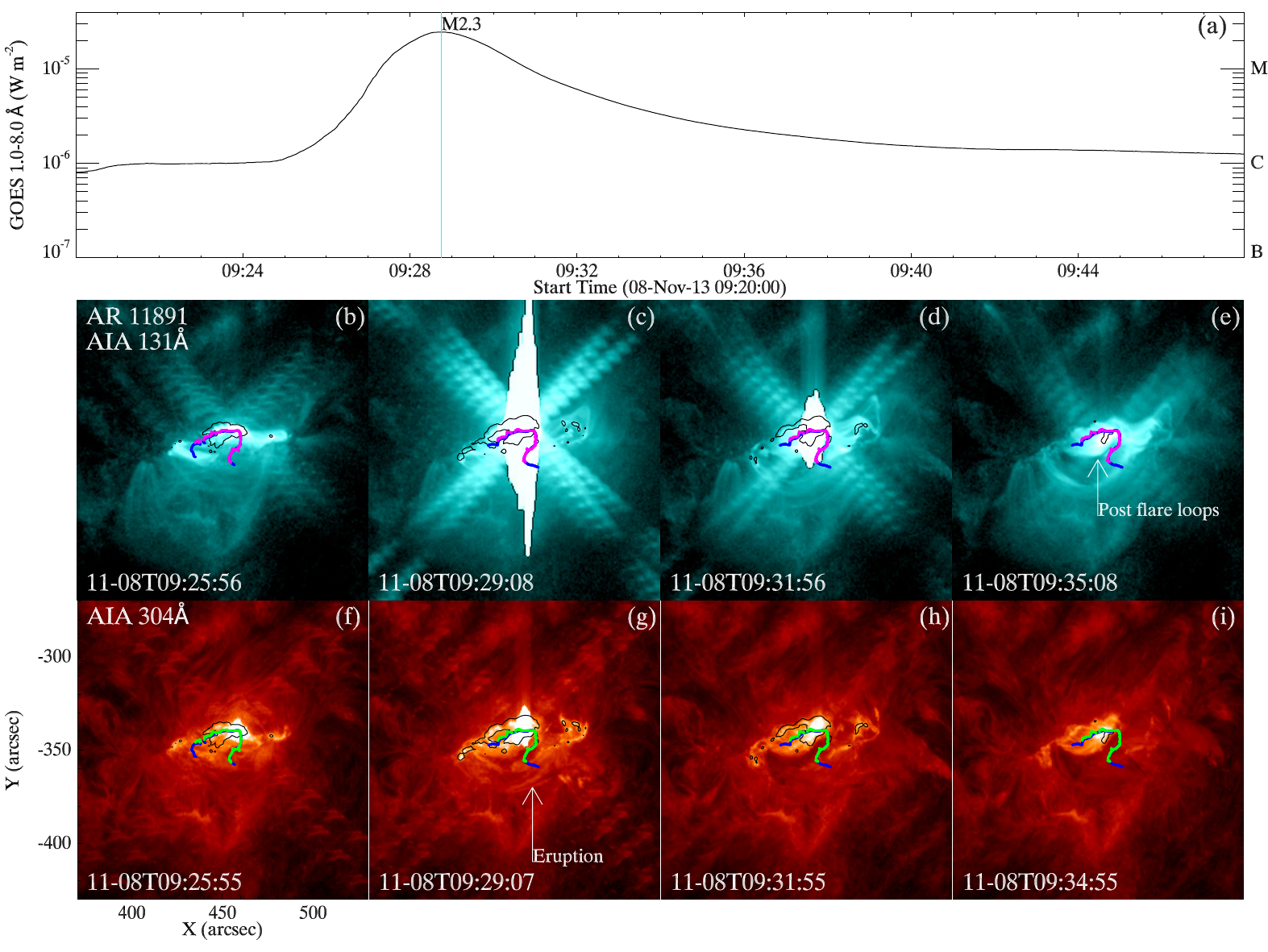}
\caption{
The first major activity occurred in NOAA AR 11891. Same layout as Figure~\ref{fig:11162_flr}. 
An animation lasting from 2013-11-08T09:25 to 2013-11-08T09:37 is available online. 
}\label{fig:11891_flr}
\end{center}
\end{figure*}

The first major eruption from this AR is an M2.3 class flare 
which starts from 2013-11-08T09:22. 
It is accompanied by a CME propagating away with a velocity of around 207 km~s$^{-1}$ (see also Table~\ref{tb:props}). 
Brightenings at the cPIL between PA and NB are observed in both the 131~\AA~passband and 304~\AA~passband, revealing that the eruption initiates from here (Figure~\ref{fig:11891_flr}). 
Moreover, mass eruption is seen in the 304~\AA~passband (Figure~\ref{fig:11891_flr}(g)).  Post-flare loops appear across the PIL in the 131~\AA~passband after the flare (Figure~\ref{fig:11891_flr}(e)). 
The flaring ribbons appear at both sides of 
the cPIL in the 1600~\AA~passband (black contours in Figure~\ref{fig:11891_flr}), confirming that the cPIL is the eruption source.

\subsubsection*{The cPIL in NOAA AR 12089}\label{susubbsec:cpil_12089}

\begin{figure*}
\begin{center}
\epsscale{1.25}
\plotone{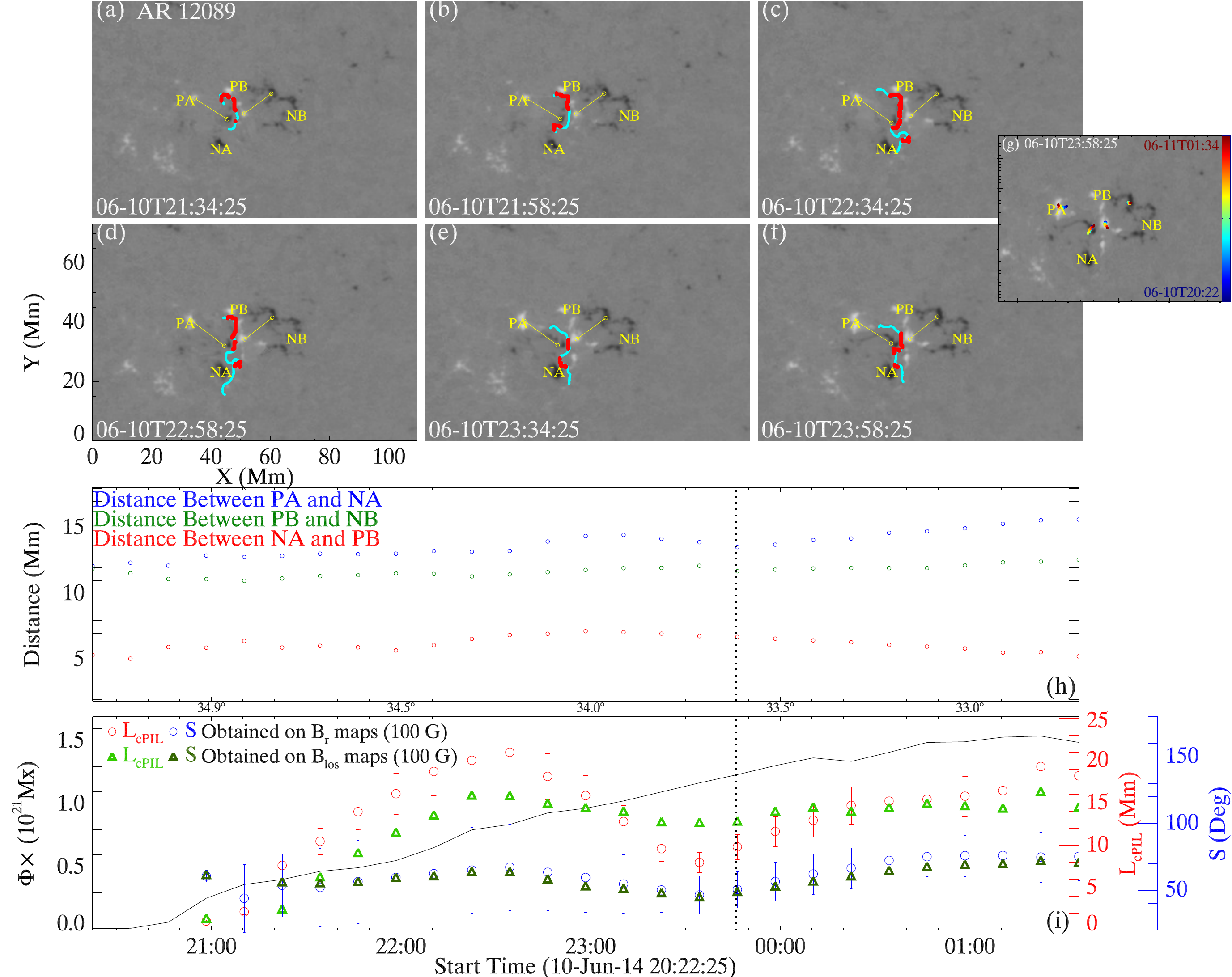}
\caption{
The evolution of NOAA AR 12089. Same layout as Figure~\ref{fig:11162_b}. 
An animation of the magnetograms lasting from 2014-06-10T20:22 to 2014-06-12T22:46 is available online. 
}\label{fig:12089_b}
\end{center}
\end{figure*}

NOAA AR 12089 is composed of two bipoles, bipole A and bipole B, which start to emerge from around 2014-06-10T20:22 simultaneously.  
Its collisional shearing case belongs to case A. 
Until the flare, the AR crosses the region between Stonyhurst longitude 31$^{\circ}$E to 28$^{\circ}$E, with the disk-centered angle changing from $35.2^{\circ}$ to $33.6^{\circ}$.  
The nonconjugated polarities NA and PB stays quite close (Figure~\ref{fig:12089_b}(a)-(f)), 
so that the PIL between them is classified as a cPIL. 
The first major activity occurs only 4 hours after the emergence onset, 
thus the conjugated polarities do not separate too much as shown by their flux-weighted centroids trajectories 
(Figure~\ref{fig:12089_b}(g)). 
The PA slightly moves to the northeast while NA moves northwestward. 
In the meantime, PB moves southward whilst NB moves westward. 
The distance between PA and NA slightly increases from 12.1~Mm to 13.5~Mm, and that between PB and NB remains 
around 11.8~Mm (Figure~\ref{fig:12089_b}(h)).   
NA and PB also has a distance remaining around 6.0~Mm, 
supporting that the two polarities are quite close from their emergence. 
Until the flare, the AR is still small, 
having unsigned magnetic flux of $1.2\times 10^{21}$~Mx (Figure~\ref{fig:12089_b} (i)). 
The length of the cPIL between NA and PB (obtained on $B_r$) increases from 1.1~Mm to 9.8~Mm, 
and the mean shear angle fluctuates slightly around 55.0$^{\circ}$. 
For $L_{cPIL}$ and $S$ obtained on $B_{los}$, the former increases from 1.5~Mm to 12.9~Mm, 
and the latter also fluctuates slightly around 55$^{\circ}$. 
The overall difference of the two sets of $L_{cPIL}$ is $-5.8\%$, 
while that of $S$ is $4.1\%$, which are all relatively small.

\begin{figure*}
\begin{center}
\epsscale{1.25}
\plotone{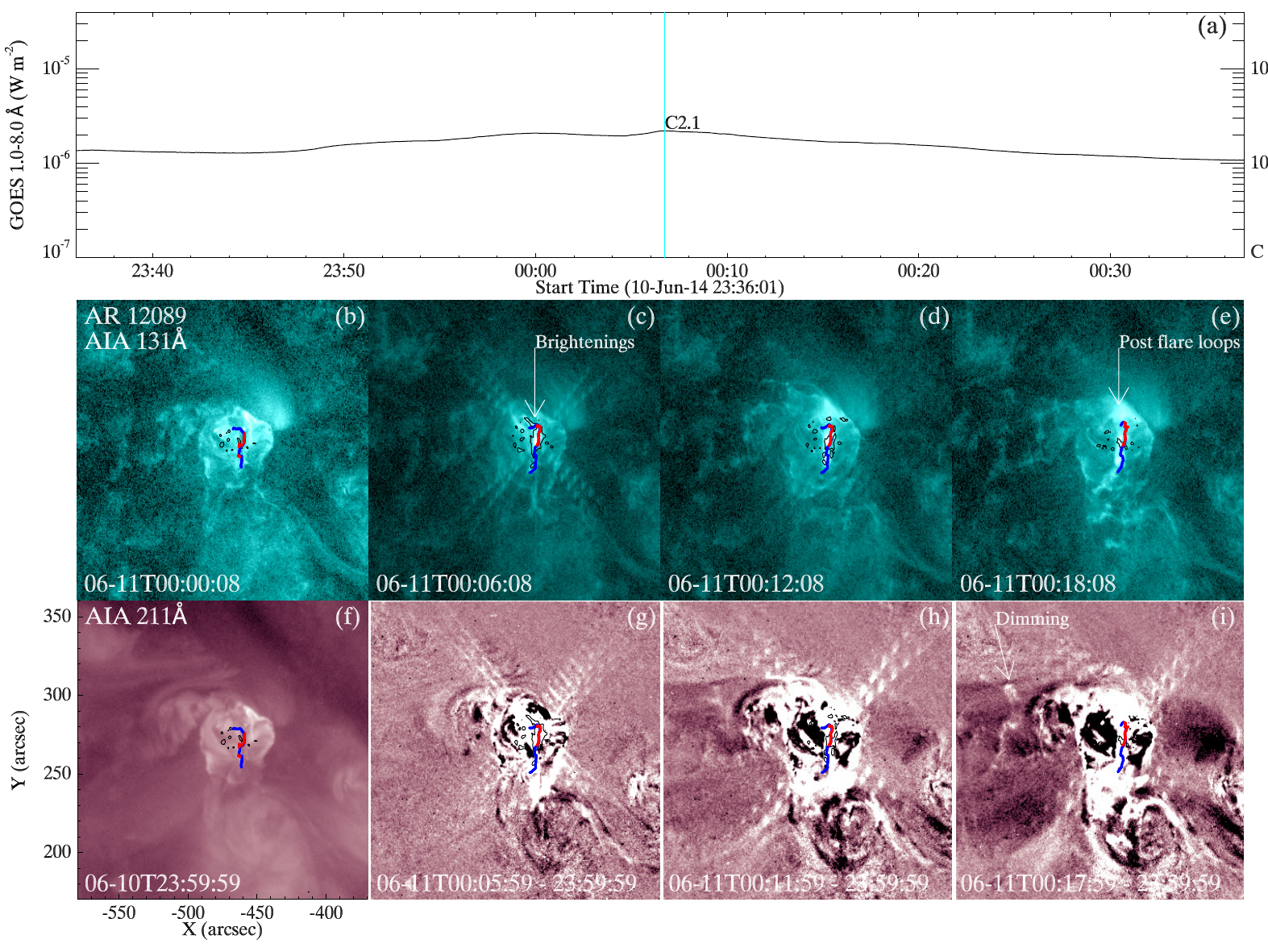}
\caption{
The first major activity occurred in NOAA AR 12089. Similar layout as Figure~\ref{fig:11162_flr}. 
Panels (f)-(i) are the base-difference images of the AIA 211~\AA~passband to show the coronal dimmings during the activity. 
The blue lines are the cPIL between PB and NA, while the red lines are the collisional parts of the PILs. 
An animation lasting from 2014-06-11T00:00 to 2014-06-11T00:23. 
is available online. 
}\label{fig:12089_flr}
\end{center}
\end{figure*}

The first major activity from the AR is a CME propagating with a velocity of around 343 km~s$^{-1}$, 
accompanied by a C2.1 class flare which starts from 2014-06-10T23:46 (Figure~\ref{fig:12089_flr}). 
In the AIA 131~\AA~passband, 
brightenings are observed occurring at the cPIL between NA and PB. 
Clear dimmings are observed near the cPIL in the AIA 211~\AA~passband, 
which indicates the mass depletion during the CME.
The flaring ribbons occurring along both sides of the cPIL in the 1600~\AA~passband confirms that it is the source PIL.

\subsection{Examples of the conjoined sPIL/cPIL}\label{subsec:s+cpil}

\subsubsection*{The conjoined sPIL/cPIL in NOAA AR 11081}\label{susubbsec:cpil_11081} 

\begin{figure*}
\begin{center}
\epsscale{1.25}
\plotone{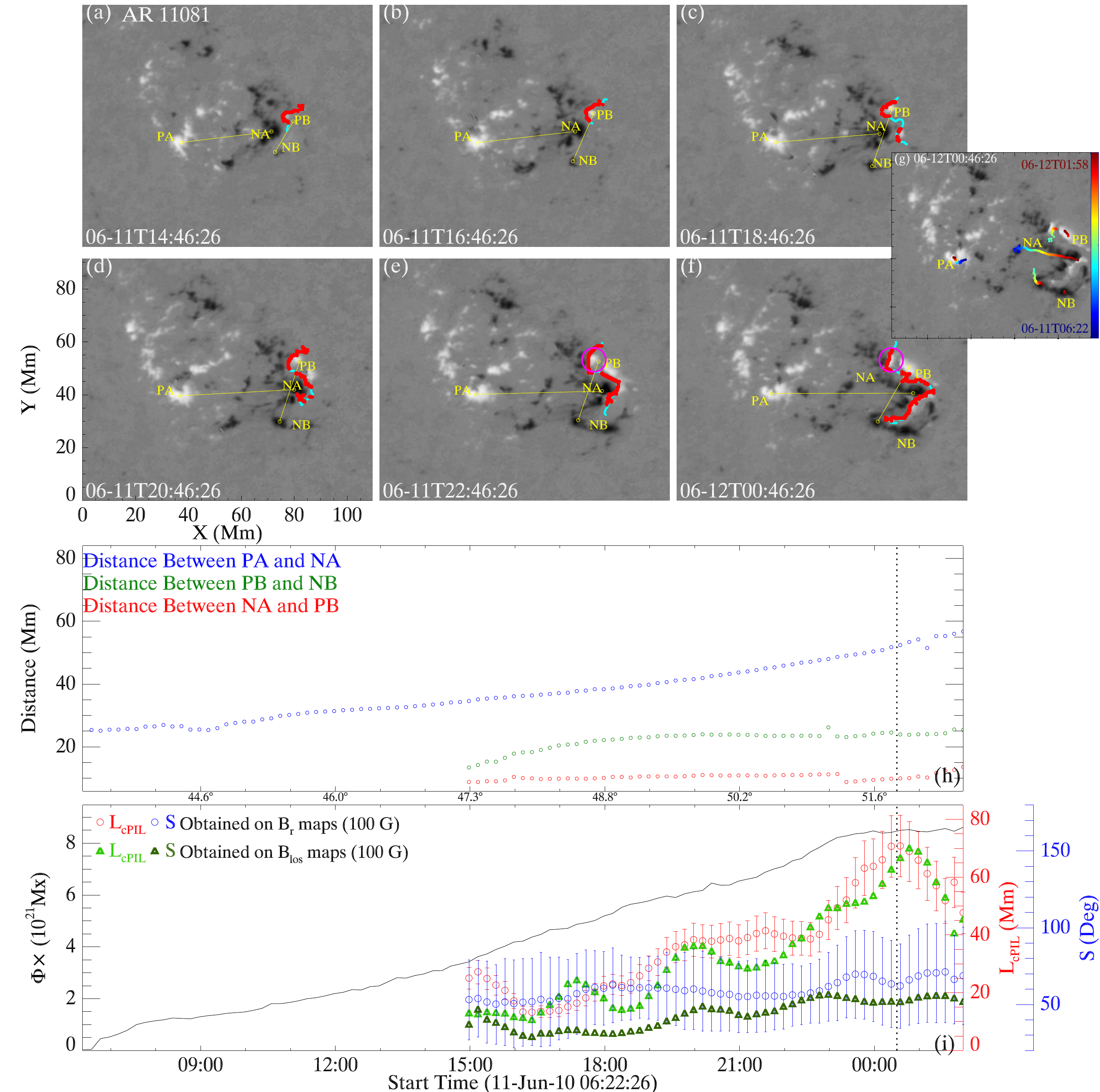}
\caption{
The evolution of NOAA AR 11081. Same layout as Figure~\ref{fig:11162_b}. 
An animation of the magnetograms lasting from 2010-06-11T06:22 to 2010-06-12T03:22 is available online. 
}\label{fig:11081_b}
\end{center}
\end{figure*}

NOAA AR 11081 presents a multi-polar configuration, which is composed of more than three bipoles emerging sequentially.  
We focus on the two lately emerged bipoles which are named as bipole A and bipole B. 
From the emergence until the flare, the AR transits from the Stonyhurst longitude 38$^{\circ}$W to 49$^{\circ}$W, with the disk-centered angle changing from $43.4^{\circ}$ to $51.9^{\circ}$.  
Bipole A starts to emerge from around 2010-06-11T06:22 (see Figure~\ref{fig:11081_b}(a)-(f)). 
About 9 hours later, bipole B starts to emerge at the west of bipole A, 
with both its polarities (PB and NB) 
located close to the negative polarity of bipole A (NA). 
As emergence goes on, NA gradually intrudes in between PB and NB. A PIL can be drawn between PB and the group of NB and NA (cyan line in Figure~\ref{fig:11081_b}), which is the active PIL of the AR. 
Collisional signatures appear on it (red line parts in Figure~\ref{fig:11081_b}). 
Shrinkage of the positive polarities is also observed 
(enclosed in magenta circles in Figure~\ref{fig:11081_b}), indicating the flux cancellation. 
Since the PIL is an integral 
system of the sPIL between PB and NB and the cPIL between PB and NA~\citep[firstly identified in][]{Chintzoglou_2019}, it is classified as a conjoined sPIL/cPIL. 
The trajectories of the flux-weighted centroids 
exhibit that PA moves northeastward while NA moves southwestward. 
PB firstly moves to the northwest then reverses toward the southwest, 
while NB moves southwestward (Figure~\ref{fig:11081_b}(g)).

Until the flare, 
the distance between PA and NA increases from 25.4~Mm to 52.4~Mm, 
and that between NB and PB increases from 13.4~Mm to 24.0~Mm, 
supporting the separation of conjugated polarities (Figure~\ref{fig:11081_b}(h)). 
The distance between NA and PB remains relatively invariant around 10~Mm, 
which indicates that the collision starts with the emergence of bipole B. The collisional shearing case thus is case B.  
The unsigned magnetic flux of the AR grows to $8.5\times 10^{21}$~Mx (Figure~\ref{fig:11081_b}(i)). 
The length of the collisional PIL part obtained on $B_r$ increases from 25.1~Mm to 70.8~Mm, 
and the mean shear angle slightly increases from 54.9$^{\circ}$ to 62.1$^{\circ}$. 
$L_{cPIL}$ and $S$ obtained on $B_{los}$ show the similar trend. The former increases from 11.1~Mm to 66.8~Mm, 
and the latter increases from 37.6$^{\circ}$ to 52.0$^{\circ}$. 
The difference of $L_{cPIL}$ from the two datasets is $-16.3\%$, and that of $S$ is $-27.1\%$, 
indicating that the values from $B_{los}$ are smaller 
compared to the ones from $B_r$ in this event.

\begin{figure*}
\begin{center}
\epsscale{1.25}
\plotone{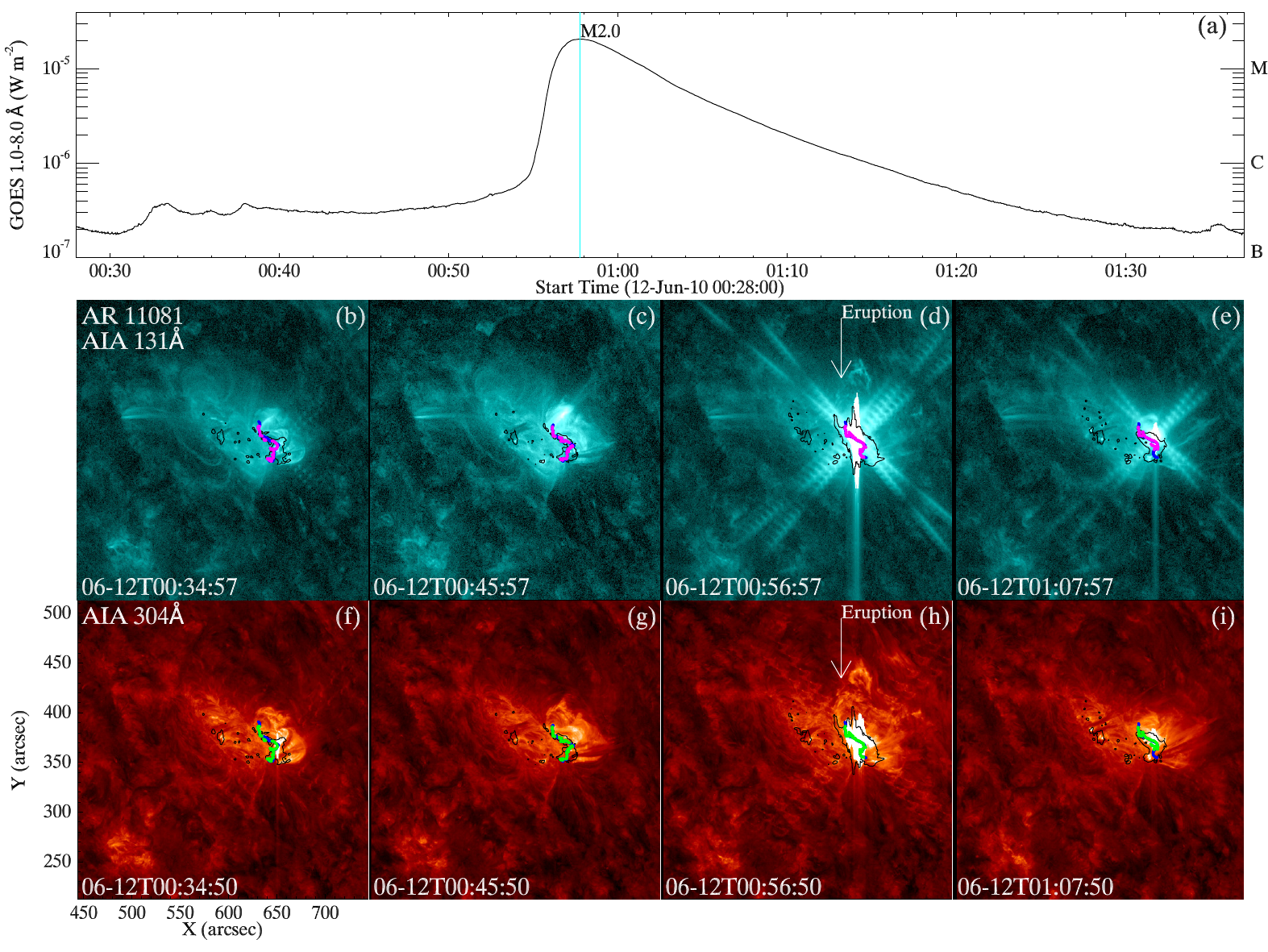}
\caption{
The first major activity occurred in NOAA AR 11081. Same layout as Figure~\ref{fig:11162_flr}. 
An animation lasting from 2010-06-12T00:35 to 2010-06-12T01:11 is available online. 
}\label{fig:11081_flr}
\end{center}
\end{figure*}

The first major activity of the AR is an M2.0 class flare starts from 2010-06-12T00:30, 
accompanied by a CME propagating away with a velocity of 486~km s$^{-1}$ (Figure~\ref{fig:11081_flr}). In both the AIA 131~\AA~and 304~\AA~passbands, brightenings and mass eruption are observed near the active PIL. The flaring ribbons in the 1600~\AA~passband appear at both sides of the PIL, confirming it is the eruption source.

\subsubsection*{The conjoined sPIL/cPIL in NOAA AR 11440}\label{susubbsec:cpil_11440} 

\begin{figure*}
\begin{center}
\epsscale{1.25}
\plotone{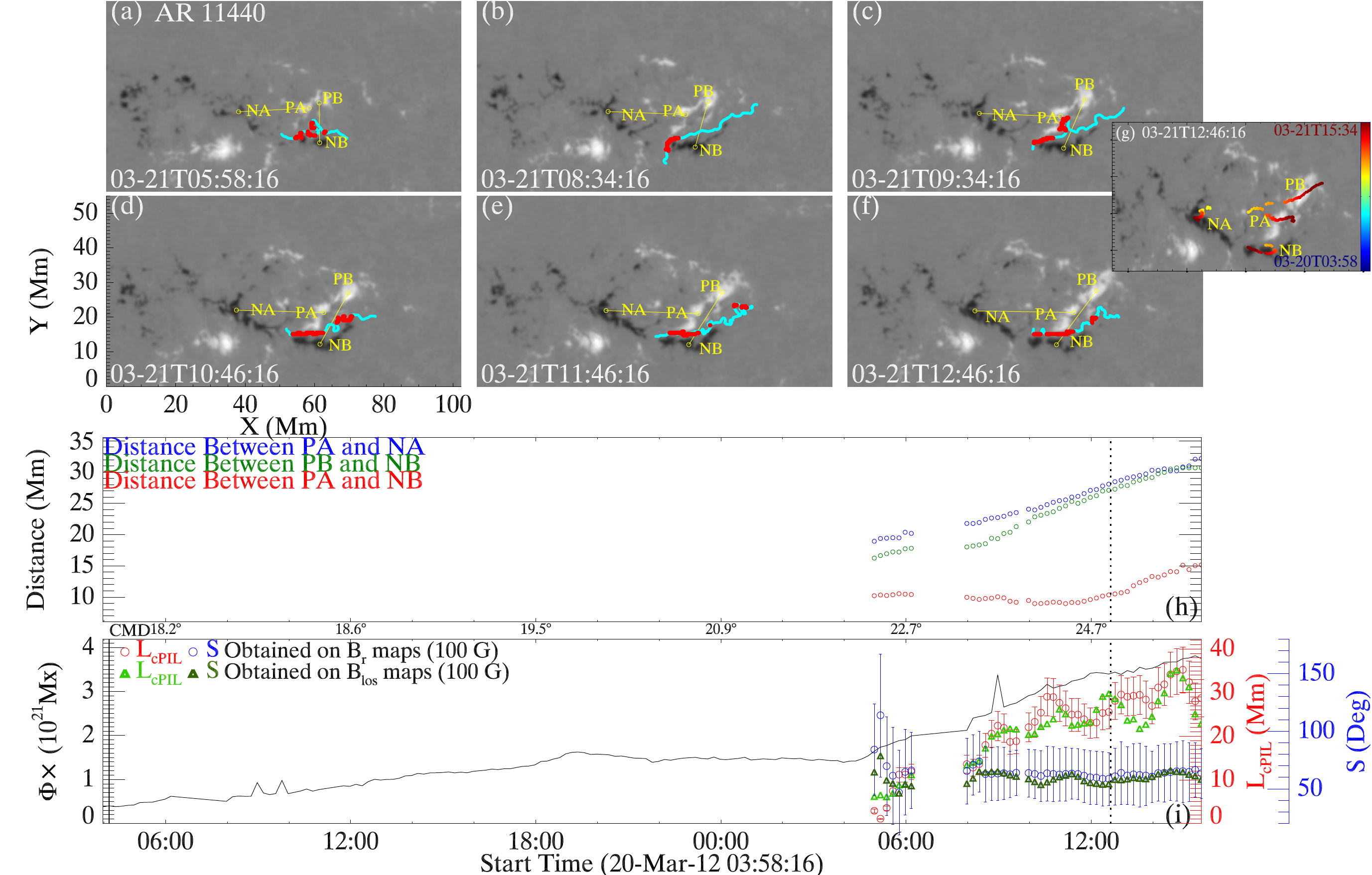}
\caption{
The evolution of NOAA AR 11440. Same layout as Figure~\ref{fig:11162_b}. 
An animation of the magnetograms lasting from 2012-03-21T02:58 to 2012-03-21T17:34:16 is available online. 
}\label{fig:11440_b}
\end{center}
\end{figure*}

NOAA AR 11440 transits from the Stonyhurst longitude 00$^{\circ}$E to 19$^{\circ}$W until the flare, with its disk-centered angle changing from $18.1^{\circ}$ to $24.9^{\circ}$. It exhibits a multi-polar configuration, 
which can be roughly divided into three bipoles (Figure~\ref{fig:11440_b}(a)-(f)). 
We focus on the two bipoles which emerge from around 2012-03-21T02:58, 
named as bipole A and bipole B. They emerge simultaneously. The collisional shearing case is case A. 
Bipole B is located at the west of bipole A, being closer to PA. As emergence goes on, PA moves in between PB and NB, forming a PIL between NB and the group of PA and PB (cyan line in Figure~\ref{fig:11440_b}). Collision signatures are observed on the PIL from the early stage of the emergence (red line part in Figure~\ref{fig:11440_b}). 
Disappearance of small polarities patches, which indicates the flux cancellation, is observed at the PIL (see the movie associated with Figure~\ref{fig:11440_b}). 
Since the PIL is an integral system of sPIL and cPIL, we classify it as a conjoined sPIL/cPIL. 
The trajectories of the flux-weighted centroids of the polarities show that (Figure~\ref{fig:11440_b}(g)), PA moves westward while NA moves southeastward; PB moves northwestward while NB moves eastward. 

Until the flare, the distance between PA and NA increases from 13.1~Mm to 28.1~Mm (Figure~\ref{fig:11440_b}(h)), 
and that between PB and NB increases from 16.2~Mm to 27.1~Mm. 
The distance between PA and NB remains almost unchanged around 10~Mm. The results prove the separation between the conjugated polarities, and quite close distance 
between PA and NB from the emergence onset. 
The AR grows into a medium sized region 
with the unsigned magnetic flux of $3.4\times 10^{21}$~Mx (Figure~\ref{fig:11440_b}(i)). The length of the collisional PIL parts obtained on $B_r$ increases from 2.9~Mm to 25.6~Mm, 
and the mean shear angle remains almost invariant around 60$^{\circ}$. 
For $B_{los}$ dataset, $L_{cPIL}$ also increases from 6.8~Mm to 29.8~Mm, while $S$ remains around 60$^{\circ}$. 
The difference of $L_{cPIL}$ and $S$ on the two datasets are $10.6\%$ and $-7.9\%$, respective, 
indicating the increase of $L_{cPIL}$ and decrease of $S$ when changing the dataset from $B_r$ to $B_{los}$ in this event.   

\begin{figure*}
\begin{center}
\epsscale{1.25}
\plotone{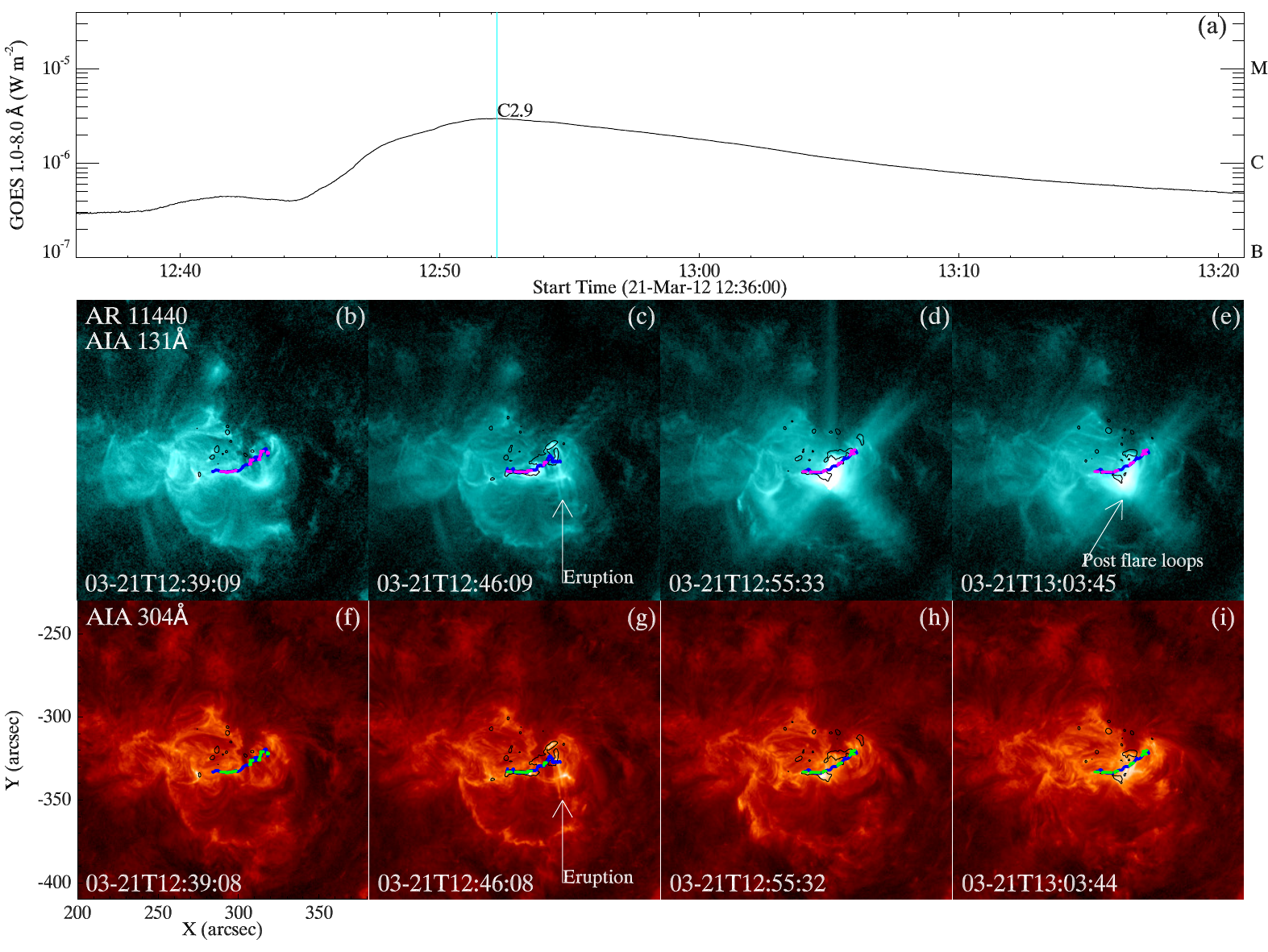}
\caption{
The first major activity occurred in NOAA AR 11440. Same layout as Figure~\ref{fig:11162_flr}. 
An animation lasting from 2012-03-21T12:39 to 2012-03-21T13:10:57 is available online. 
}\label{fig:11440_flr}
\end{center}
\end{figure*}

The first major activity from the AR is a slow CME with a velocity of 387~km s$^{-1}$, associated with a C2.9 class flare starting from 2012-03-21T12:38 (Figure~\ref{fig:11440_flr}). 
In both AIA~131~\AA~and 304~\AA~passbands, mass eruption is observed at the source PIL. Flaring ribbons along both sides of the PIL are observed in the AIA~1600~\AA~passband (black contours in Figure~\ref{fig:11440_flr}), confirming it is the eruption source.

\subsubsection*{The conjoined sPIL/cPIL in NOAA AR 11776}\label{susubbsec:cpil_11776} 

\begin{figure*}
\begin{center}
\epsscale{1.25}
\plotone{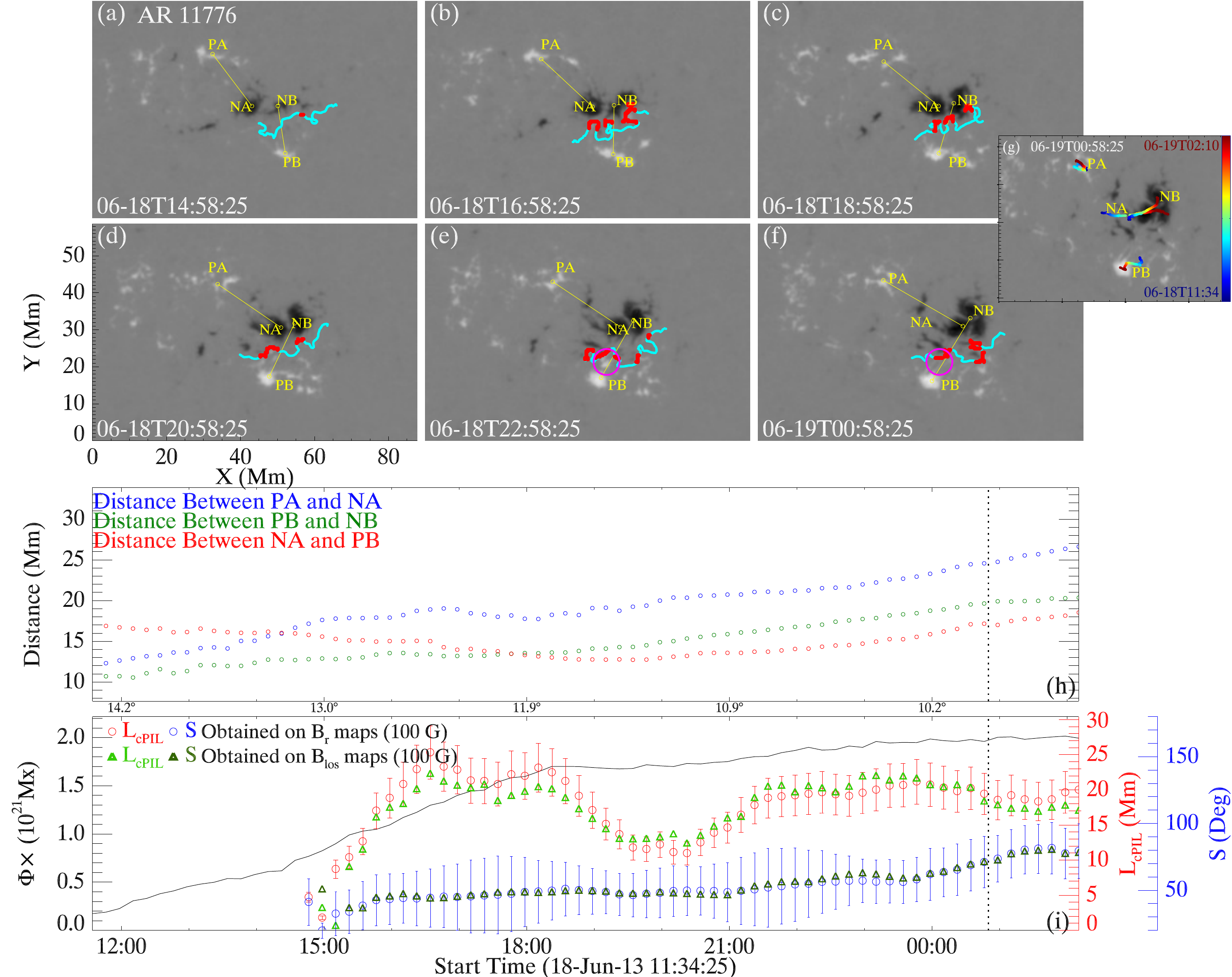}
\caption{
The evolution of NOAA AR 11776. Same layout as Figure~\ref{fig:11162_b}. 
An animation of the magnetograms lasting from 2013-06-18T11:34 to 2013-06-19T03:46 is available online. 
}\label{fig:11776_b}
\end{center}
\end{figure*}

NOAA AR 11776 transits from the Stonyhurst longitude 11$^{\circ}$E to 03$^{\circ}$E until the flare, 
with its disk-centered angle changing from $14.4^{\circ}$ to $10.0^{\circ}$. 
It is composed of two bipoles, bipole A and bipole B, which starts to emerge from around 2013-06-18T07:12 (Figure~\ref{fig:11776_b}) simultaneously. 
The collisional shearing case is case A. 
The negative polarities of the two bipoles reside close to each other. 
As emergence goes on, 
the conjugated polarities of each bipole separate, 
the two negative polarities collide and coalesce with each other. Consequently, no distinguished boundary between them could be drawn on the magnetograms.  
The PIL formed between PB and the group of NA and NB is the active PIL of the AR. On the PIL, collisional signatures appear and grow gradually. It is thus classified as a conjoined sPIL/cPIL. 
Disappearance of the positive polarities is observed 
(enclosed in magenta circles in Figure~\ref{fig:11776_b}), 
indicating the flux cancellation. From the trajectories of the flux-weighted centroids of the polarities (Figure~\ref{fig:11776_b}(g)) one can see, 
PA firstly moves westward and then turns to the northeast. 
PB moves eastward. Both NA and NB roughly move toward the west, with NB slightly inclining to the north and NA slightly inclining to the south. 

Until the flare, 
the distance between PA and NA increases from 12.3~Mm to 24.6~Mm, 
and that of PB and NB increases from 10.7~Mm to 19.6~Mm (Figure~\ref{fig:11776_b}(h)). 
The distance between NA and PB changes a little, slightly decreasing from 16.9~Mm to 12.7~Mm, 
then increasing to 17.2~Mm again. The results confirm the separation between the conjugated polarities, and the collision between PB and the group of NA and NB. 
The AR is still small with the unsigned magnetic flux of $2.0\times 10^{21}$~Mx (Figure~\ref{fig:11776_b}(i)). 
The length of the collisional PIL part obtained on $B_r$ increases from 4.7~Mm to 19.5~Mm, and the mean shear angle increases from 41.3$^{\circ}$ to 71.2$^{\circ}$. 
For the $B_{los}$ dataset, $L_{cPIL}$ also increases from 3.6~Mm to 18.4~Mm, while $S$ increases from 47.4$^{\circ}$ to 71.6$^{\circ}$. 
The difference of $L_{cPIL}$ on the two datasets is $-4.4\%$, and that for $S$ is $4.8\%$, which both are quite small.

\begin{figure*}
\begin{center}
\epsscale{1.25}
\plotone{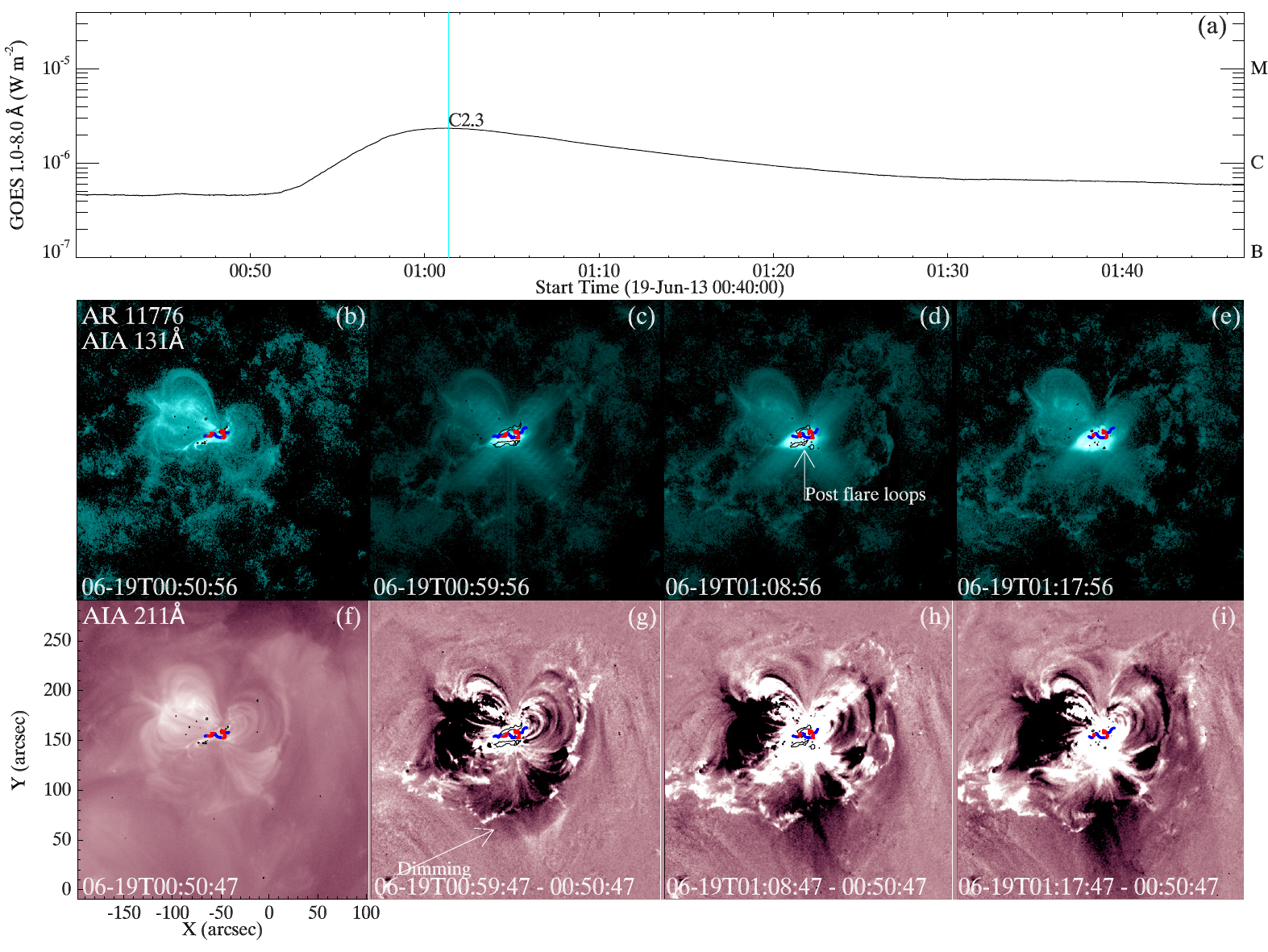}
\caption{
The first major activity occurred in NOAA AR 11776. Same layout as Figure~\ref{fig:12089_flr}. 
An animation lasting from 2013-06-19T00:51 to 2013-06-19T01:21 is available online. 
}\label{fig:11776_flr}
\end{center}
\end{figure*}

The first major activity from AR 11776 is a CME accompanied by a C2.3 class flare (Figure~\ref{fig:11776_flr}). 
The CME runs out with a velocity of around 287~km s$^{-1}$, 
with the associated flare starting from around 2013-06-19T00:50 (see Table~\ref{tb:props}). 
During the flare, brightenings are observed at the PIL between PB and the group of NA and NB in both the AIA 131~\AA~and the 211~\AA~passbands. Moreover, post-flare loops appear along the PIL (Figure~\ref{fig:11776_flr}(d)). 
Dimmings are observed in the AIA 211~\AA~passband, verifying the depletion of the mass during the CME.   
The flaring ribbons in the 1600~\AA~passband (black contours in Figure~\ref{fig:11776_flr}) are observed along the PIL, confirming that it is the eruption source.

\subsection{Examples of the conjoined sPIL/ePIL}\label{subsec:s+epil} 

\subsubsection*{The conjoined sPIL/ePIL in NOAA AR 11422}\label{susubbsec:cpil_11422} 

\begin{figure*}
\begin{center}
\epsscale{1.25}
\plotone{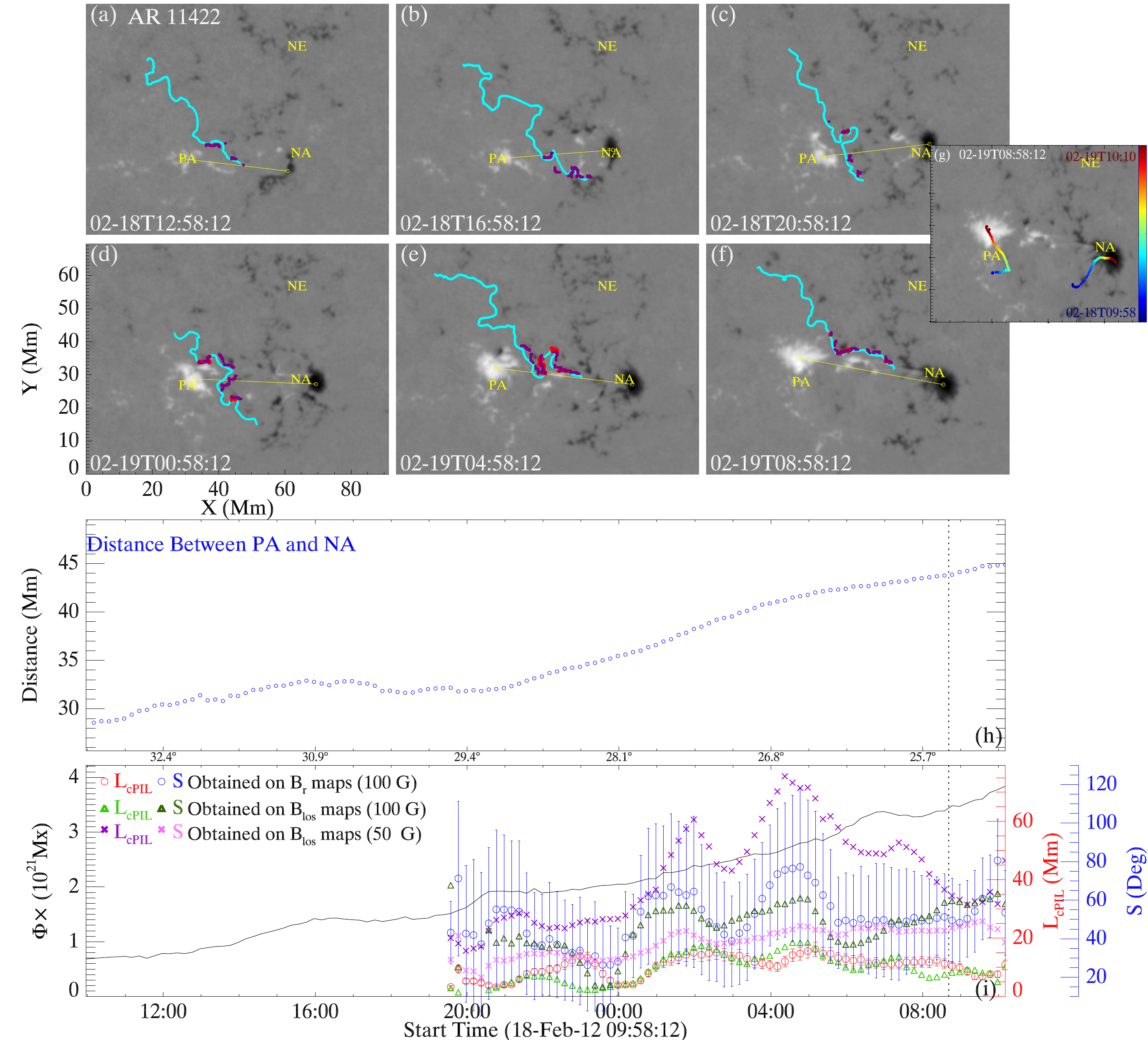}
\caption{
The evolution of NOAA AR 11422. Similar layout as Figure~\ref{fig:11162_b}. NE denotes the external negative polarity. The collisional PIL part detected at the threshold of 50~G on $B_{los}$ dataset is overplotted on the $B_r$ magnetograms for comparison (in purple color). Its 
$L_{cPIL}$ and $S$ 
is also shown in panel (i).  
There is slight inconsistency between the spatial locations of the purple line and the cyan line (or red line), which may result from the difference between the corrected $B_{los}$ and the $B_r$.  
An animation of the magnetograms lasting from 2012-02-18T09:58 to 2012-02-19T11:22 is available online. 
}\label{fig:11422_b}
\end{center}
\end{figure*}

NOAA AR 11422 is a bipolar region which starts to emerge from around 2012-02-18T09:58 (Figure~\ref{fig:11422_b}(a)-(f)). It transits from the Stonyhurst longitude 24$^{\circ}$E to 11$^{\circ}$E until the flare, with the disk-centered angle changing from $33.2^{\circ}$ to $25.5^{\circ}$.  
An external negative polarity patch (NE) is preexisting at the north of the bipole.  
The PIL formed between the positive polarity PA and the negative polarities NA and NE is the active PIL of the AR. 
Collisional signatures are also observed at the PIL. 
The collisional PIL part detected on $B_r$ series at 100~G (red line parts in Figure~\ref{fig:11422_b}) seems not that significant compared to the entire, longer, PIL (cyan lines in Figure~\ref{fig:11422_b}). 
Note that the strength of the magnetic field near the two conjoined sPIL/ePIL 
is lower compared to the others, therefore the large thresholds may underestimate their collision. 
We thus also display the collisional PIL parts detected at 50~G on $B_{los}$ 
for the sPIL/ePIL case for comparison. 
It is seen that the collision detected at lower threshold (purple line parts in Figure~\ref{fig:11422_b}(a)-(f)) is apparently longer 
than the one detected at higher thresholds (red line parts), mostly occurring between PA and the external NE.  
Since the active PIL contains self part between PA and NA and external part between PA and NE, 
it is classified as a conjoined sPIL/ePIL. 

The trajectories of the flux-weighted centroids of the polarities show that PA moves northeastward while NA moves northwestward, separating from each other (Figure~\ref{fig:11422_b}(g)), which is confirmed by the evolution of the distance between them (Figure~\ref{fig:11422_b}(h)). 
From the start of the flux emergence until the flare, 
the distance between the polarities increases from 28.6~Mm to 43.8~Mm. 
The unsigned  magnetic flux of the AR grows to $3.5\times 10^{21}$~Mx (Figure~\ref{fig:11422_b}(i)). The length and the mean shear angle of the collisional PIL part obtained at 100~G on $B_r$ remain relatively small. The former increases from 3.3 to 10.5~Mm, and the latter slightly increases from 43.2$^{\circ}$ to 49.4$^{\circ}$.  
The parameters obtained on $B_{los}$ at 100~G show the similar trend, 
with $L_{cPIL}$ increasing 
from 1.1~Mm to 8.2~Mm, and $S$ increasing 
from 13.8$^{\circ}$ to 59.7$^{\circ}$. 
The difference of $L_{cPIL}$ on the two datasets is $-8.7\%$, and that of $S$ is $-9.0\%$. 
When changing the detection threshold to 50~G, the $L_{cPIL}$ is significantly longer (Figure~\ref{fig:11422_b}(i)), increasing from 3.3~Mm to 34.5~Mm. The shear angle also increases from 37.2$^{\circ}$ to 45.7$^{\circ}$. The results indicate that the collision in the conjoined sPIL/ePIL in AR 11422 is not trivial.

\begin{figure*}
\begin{center}
\epsscale{1.25}
\plotone{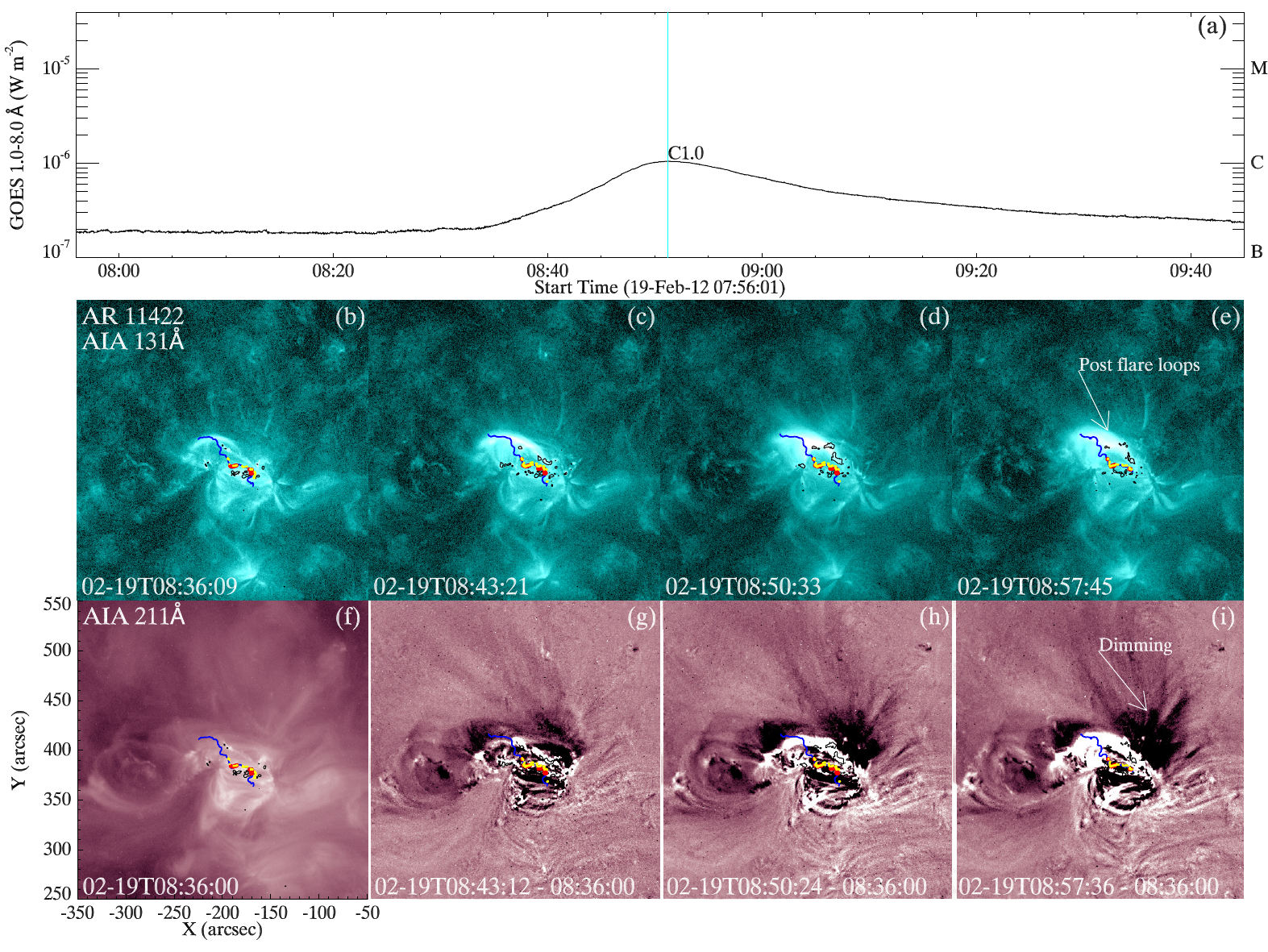}
\caption{
The first major activity occurred in NOAA AR 11422. Similar layout as Figure~\ref{fig:11776_flr}. 
The collisional PIL part detected at the threshold of 50~G on $B_{los}$ dataset is also overplotted for comparison (in yellow color).  
An animation lasting from 2012-02-19T08:36 to 2012-02-19T09:09 is available online. 
}\label{fig:11422_flr}
\end{center}
\end{figure*}

The first major activity from the AR is a slow CME propagating with a velocity of around 238 km~s$^{-1}$, 
accompanied by a C1.0 class flare which starts from around 2012-02-19T08:41 (See Table~\ref{tb:props}). 
The images at both the AIA~131~\AA~and 211~\AA~passbands clear show that the eruption generates brightenings along the active PIL 
(Figure~\ref{fig:11422_flr}). 
Post-flare loops across the PIL are also observed (Figure~\ref{fig:11422_flr}(e)). Furthermore, regions of dimming appear 
in the 211~\AA~passband, indicating the mass depletion associated with the CME.  
Flaring ribbons in the AIA 1600~\AA~passband also appear along the PIL (black contours in Figure~\ref{fig:11422_flr}), confirming that it is the eruption source.

\subsubsection*{The conjoined sPIL/ePIL in NOAA AR 11870}\label{susubbsec:cpil_11870} 

NOAA AR 11870 transits from the Stonyhurst longitude 14$^\circ$E to 28$^\circ$W until the flare, with the disk-centered angle changing from $24.1^{\circ}$ to $32.5^{\circ}$.  
It is a bipolar region which starts to emerge from around 2013-10-13T06:10 (Figure~\ref{fig:11870_b}(a)-(f)). A patch of dispersive negative polarity is preexisting at its south (NE). 
The PIL between PA and the group of NA and NE is the active PIL of the AR (cyan line in Figure~\ref{fig:11870_b}(a)-(f)), which is obvious a conjoined sPIL/ePIL. 
Collisional signatures also appear on the PIL. 
The one detected at higher thresholds of 100~G (on $B_r$, red line parts in Figure~\ref{fig:11870_b}(a)-(f)) is quite short compared to the entire PIL. 
The one detected at lower thresholds of 50~G (on $B_{los}$, purple line parts) is longer. 
It is seen that in the early emergence stage (Figure~\ref{fig:11870_b}(a)-(c)), quite a part of 
the collision occurs between the conjugated PA and NA, 
while near to the flare (Figure~\ref{fig:11870_b}(d)-(f), the most of the collision occurs between PA and the external NE.

As emergence goes on, the two conjugated polarities separate (Figure~\ref{fig:11870_b}(g)). 
The flux-weighted centroids of the polarities change several times. In the rough, PA moves southwestward, while NA moves to the north at first then turns to southeast. Until the flare,  
the distance between PA and NA increases from 25.0~Mm to 53.4~Mm (Figure~\ref{fig:11422_b}(h)). 
The unsigned magnetic flux of the AR grows to $5.1\times 10^{21}$~Mx (Figure~\ref{fig:11870_b}(i)). The length of the collisional PIL part obtained on $B_r$ at 100~G evolves dramatically, increasing from 0.7~Mm to 28.0~Mm, then decreasing to 3.9~Mm. 
The mean shear angle increases from 63.0$^\circ$ to 84.5$^\circ$. 
The parameters obtained on $B_{los}$ at 100~G show the similar evolution trend. The $L_{cPIL}$ firstly increases from 0.7~Mm to 22.8~Mm, then decreases to 2.5~Mm until the flare. The $S$ increases from 53.8$^\circ$ to 99.7$^\circ$. 
The difference of $L_{cPIL}$ on the two datasets is $-12.3\%$, and that of $S$ is $-1.5\%$.  
For the collisional PIL part detected at 50~G on $B_{los}$,  
the $L_{cPIL}$ also evolves dramatically, increasing from 2.2~Mm to as high as 61.5~Mm, then decreasing to 9.5~Mm until the flare, similar as the one obtained at the higher threshold. 
The $S$ increases from around 60.0$^\circ$ to 82.5$^\circ$. 
This indicates that the collision at this sPIL/ePIL is severe in the early emergence phase,  
but becomes mild right before the flare. 

\begin{figure*}
\begin{center}
\epsscale{1.25}
\plotone{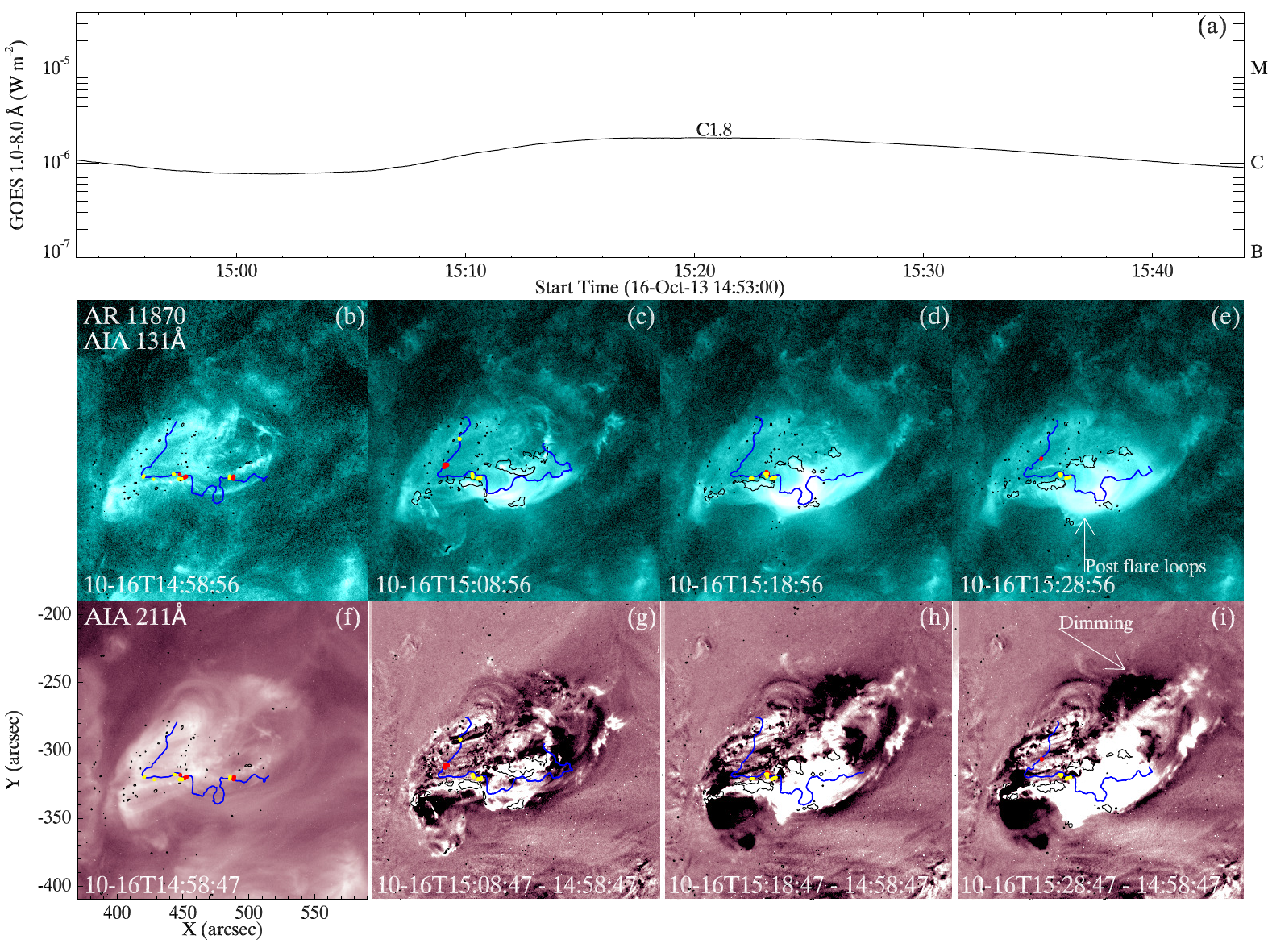}
\caption{
The first major activity occurred in NOAA AR 11870. Same layout as Figure~\ref{fig:11776_flr}. 
An animation lasting from 2013-10-16T14:58 to 2013-10-16T15:43 is available online. 
}\label{fig:11870_flr}
\end{center}
\end{figure*}

The first major activity from the AR is a CME with a velocity of 514 km~s$^{-1}$, accompanied by a C1.8 class flare starting from 2013-10-16T15:03 (Figure~\ref{fig:11870_flr}). 
In both the AIA 131~\AA~and the 211~\AA~passbands, brightenings occur along the PIL. After the flare, post-flare loops across the PIL are observed (Figure~\ref{fig:11870_flr}(e)). Dimming regions appear in the 211~\AA~passband, which indicate the mass depletion during the CME. The flaring ribbons in the 1600~\AA~passband occur along both sides of the PIL, confirming that it is the eruption source.

\subsection{Correlation between the intensity of the activities and $\overline{L}_{cPIL}$ and $\overline{S}$}
\label{subsec:stat}

We further inspect the properties of all active PILs and their correlation to the intensity of the activities. 
Specifically, we check the correlation between GOES 1-8~\AA~peak flux (F) and the mean length ($\overline{L}_{cPIL}$) and mean shear angle ($\overline{S}$) of the collisional PIL, which are averaged in a given duration prior to the activities. 
Three durations, 1 hours, 3 hours, and 5 hours are used to do the average. 
The analysis is performed at various thresholds on the two datasets to avoid the bias (see also Section~\ref{sec:obs_ana}). 
 
\begin{figure*}
\begin{center}
\epsscale{1.1}
\plotone{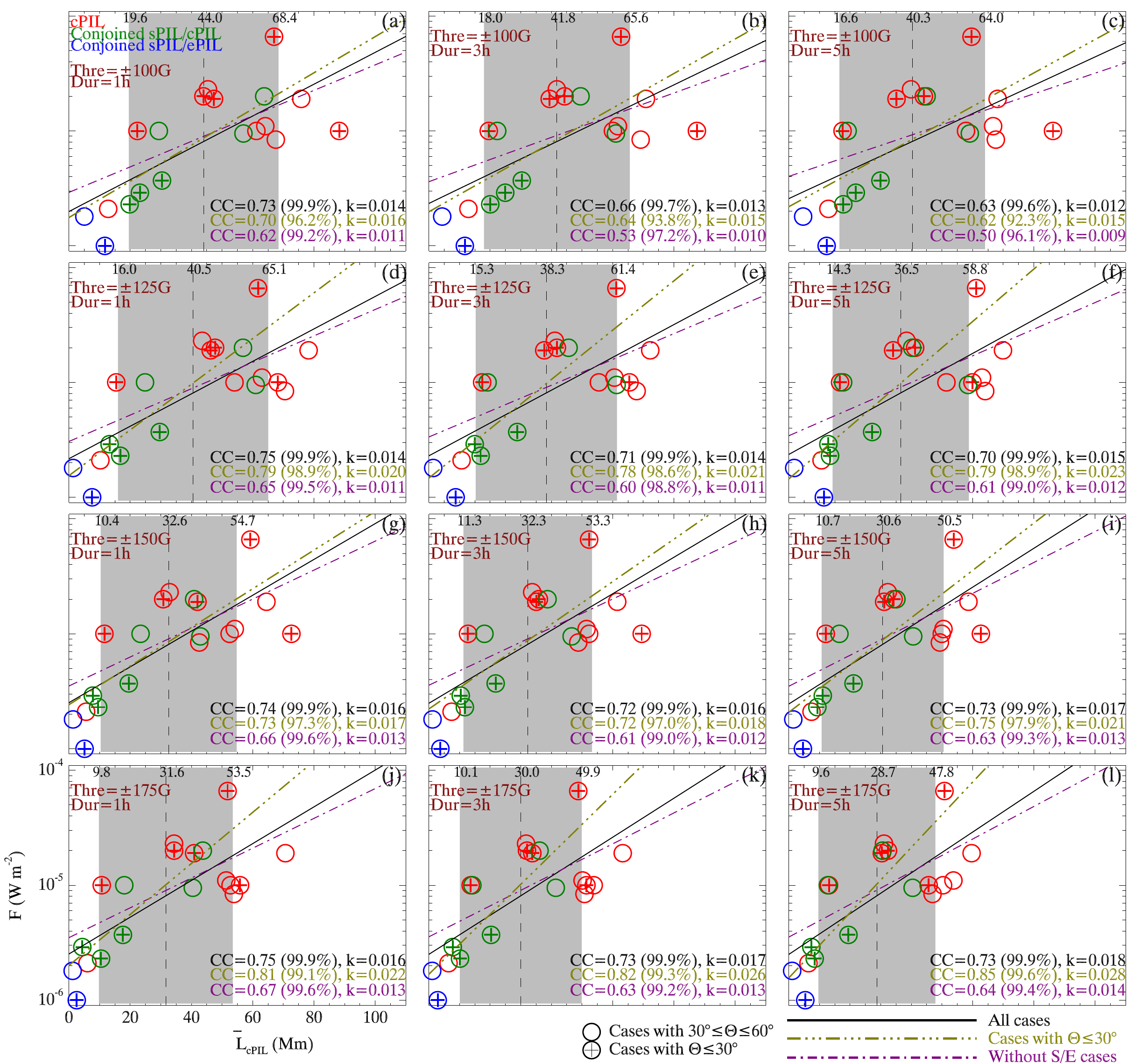}
\caption{
The scatter plots of the mean length of the collisional part of the PIL ($\overline{L}_{cPIL}$) obtained from $B_{r}$ dataset on various thresholds and averaging durations vs. the GOES 1-8~\AA~flux (F). 
For each AR, $\overline{L}_{cPIL}$ is averaged  in the given durations prior to its first major activity. 
The term ``Thre'' indicates the threshold of the cPIL detection. 
``Dur'' indicates the duration prior to the activities within which to do the average. 
The red circles show the parameters of the cPILs. 
The green ones are for the conjoined sPIL/cPIL 
and the blue ones are for the conjoined sPIL/ePIL, 
respective. The cases having $\Theta\leq 30^{\circ}$ are further marked by the ``$+$'' sign in circles. 
The black solid lines show the result of a linear fitting between $\overline{L}_{cPIL}$ and the logarithmic GOES flux for all cases, 
with the Pearson correlation coefficient CC and corresponding confidence level, and the slope $k$ shown at the bottom right. 
The purple dash-dotted lines show the linear fitting results on the sample excluding the two conjoined sPIL/ePIL cases.
The olive dash-dotted lines show the linear fitting results on the cases with $\Theta\leq 30^{\circ}$.  
The vertical dashed line in (a) indicates the average of all $\overline{L}_{cPIL}$, with shaded region covering between one standard deviation (of all $\overline{L}_{cPIL}$) below and over the average. The value of the average and the range of the shaded region are shown at the upper axis. The vertical dashed lines and the shaded regions in the other panels have the same meaning. 
}\label{fig:clshr_all_br}
\end{center}
\end{figure*}

Figure~\ref{fig:clshr_all_br} displays the scatter diagrams between F and $\overline{L}_{cPIL}$ obtained on $B_r$ series.  
It is seen that in general, the $\overline{L}_{cPIL}$ of the collisional parts of the PILs detected at lower threshold are longer than the ones detected at higher threshold. 
The values calculated from 100~G and 125~G are close, with the former being around 4~Mm longer than the latter. 
The ones calculated from 150~G are around 7~Mm shorter than the ones from 125~G, but are close (around 2~Mm higher) to the ones from 175~G. 
For a fixed threshold, the $\overline{L}_{cPIL}$ computed over different durations shows no significant difference, 
although the value slightly decreases as the duration increases. 
Overall, 
there is a relatively nontrivial correlation between the magnitude of the flares and the length of the collisional parts of the PILs, 
being established for all thresholds and durations. 
The Pearson correlation coefficients are not small, ranging from 0.63 to 0.75 with the confidence levels all greater than 99$\%$. This indicates that intenser activities tend to originate from the longer collisional PILs.

We choose the values calculated at the threshold of 100~G and duration of 3~hours for interpretation 
(Figure~\ref{fig:clshr_all_br}(b)).  
A linear fitting to the scatter plot gives a relation of $Log(F)=0.013\times\overline{L}_{cPIL}-5.63$ (the black solid line). 
The $\overline{L}_{cPIL}$ of different ARs are rather scattered. 
The collisional parts of the 2 conjoined sPIL/ePIL are the shortest,  
measured as 4.3~Mm and 11.7~Mm. Those of the 6 conjoined sPIL/cPIL are overall larger, ranging from 20.0~Mm to 61.0~Mm. The $\overline{L}_{cPIL}$ of the 11 cPILs are relatively largest, ranging from 12.9~Mm to 87.5~Mm.  
In general, the average of all $\overline{L}_{cPIL}$ is 41.8~Mm (the vertical dashed line in Figure~\ref{fig:clshr_all_br}(b)), and 68$\%$ (13$/$19) of the $\overline{L}_{cPIL}$ fall in the range of 18~Mm to 65.6~Mm (the shaded region in Figure~\ref{fig:clshr_all_br}(b)), 
which are one standard deviation ($\sigma$) below and over the average. 
Considering that F is positively correlated to $\overline{L}_{cPIL}$, 
taking the averaged lower cutoff value for $\overline{L}_{cPIL}$ (18~Mm here) as a reference value 
may be useful to assess the significance of the collision, thus to evaluate the productivity of the AR.   
It is seen that except the two conjoined sPIL/ePIL and one cPIL in AR 12089, the $84\%$ (16$/$19) of the $\overline{L}_{cPIL}$ are above the reference value.   

Note that all of the activities occur within the region of $\Theta \leq 60^{\circ}$ 
(see Table~\ref{tb:props}), with 9 cases close to the disk center (having $\Theta \leq 30^{\circ}$, marked by ``$+$'' symbol in circle in Figure~\ref{fig:clshr_all_br}). As discussed in Section~\ref{sec:obs_ana}, the criterion of $\Theta \leq 60^{\circ}$ could ensure relatively high signal-to-noise ratio of the data, 
but is still less strict than $\Theta \leq 30^{\circ}$ since either dateset suffer from the least uncertainties near the center.  
We thus further perform a correlation analysis on the 9 near center cases particularly (shown in olive color in Figure~\ref{fig:clshr_all_br}). It is seen that there is still relatively nontrivial correlation between F and $\overline{L}_{cPIL}$. The correlation coefficients range from 0.62 to 0.85, being very close to the ones obtained from the full sample at the thresholds of 100~G and 150~G, and slightly higher at the thresholds of 125~G and 175~G. The confidence levels of the correlations are lower than the ones from the full sample, ranging from $92.3\%$ to $99.6\%$, which may be resulted from the smaller sample size of 9 cases. In a word, the trend that intenser activities tend to originate from the PILs with longer collision still exists 
when only considering the cases having lower noise data.

One may also question that the relatively nontrivial correlations may be largely determined by the two conjoined sPIL/ePIL cases, 
since they have the shortest $\overline{L}_{cPIL}$ as well as the smallest flare. 
In order to check this possibility, 
we further analyze the correlation on a sample excluding the two cases (shown in purple color in Figure~\ref{fig:clshr_all_br}).
It is seen that the Pearson correlation coefficients are smaller than the ones from the full sample, decreasing by around 0.1, but are still relatively nontrivial, ranging from 0.5 to 0.67.  
The decrease of the coefficients 
suggests that the two sPIL/ePIL cases do bias the correlation. 
Moreover, since there are only two sPIL/ePIL cases, we are not able to make precise conclusion about how much this kind of case can affect the correlation. 
We can only see that the correlation coefficients are not small even after excluding the two cases, still indicating the trend that intenser activities tend to originate from PILs with longer collisional length.  

\begin{figure*}
\begin{center}
\epsscale{1.1}
\plotone{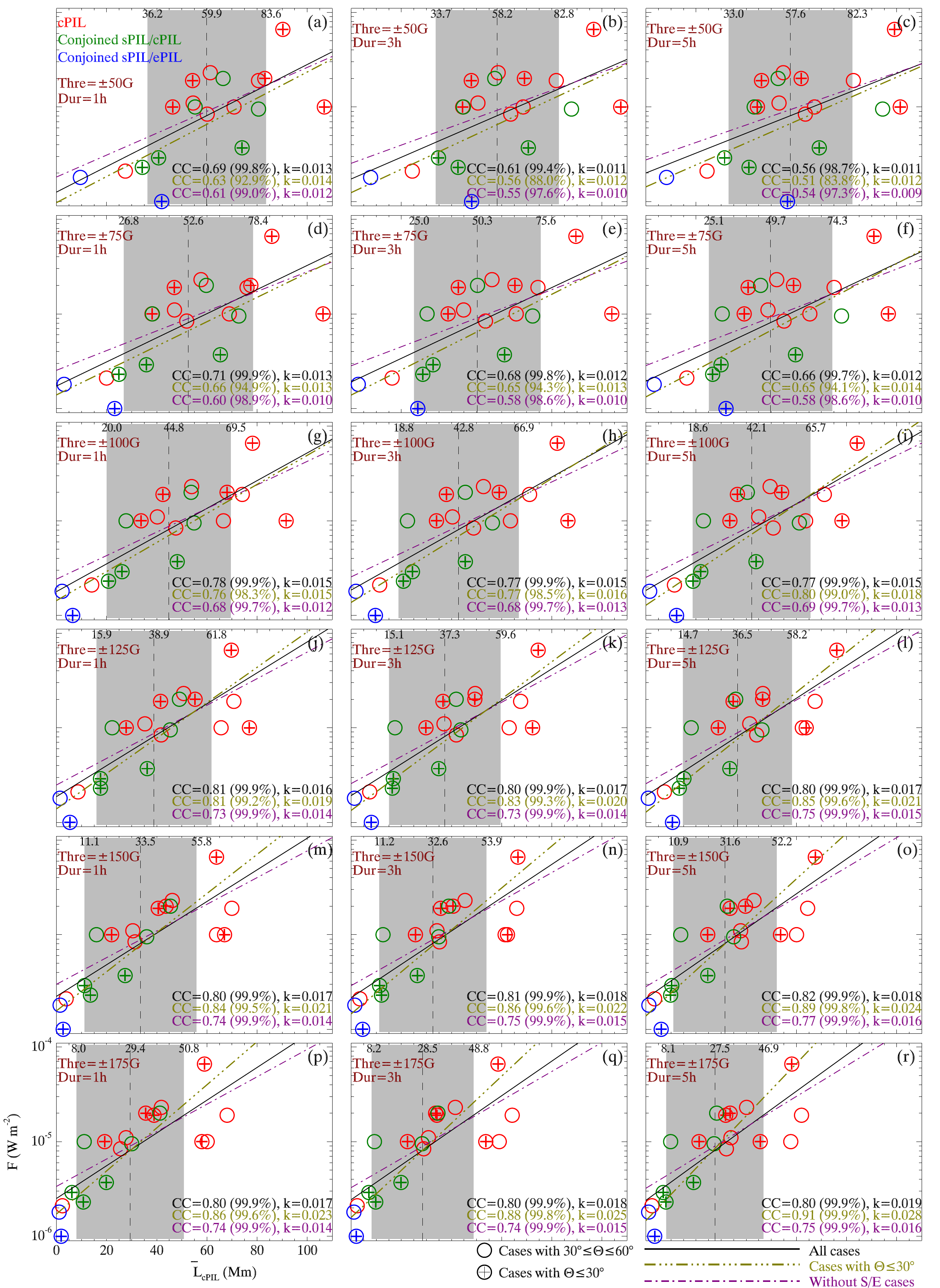}
\caption{
The scatter plots of the mean length of the collisional part of the PIL ($\overline{L}_{cPIL}$) obtained from $B_{los}$ dataset vs. the GOES 1-8~\AA~flux (F) on various thresholds and averaging durations. Similar layout as Figure~\ref{fig:clshr_all_br}.   
}\label{fig:clshr_all_los}
\end{center}
\end{figure*}

The $\overline{L}_{cPIL}$ obtained on $B_{los}$ dataset show the similar distribution (Figure~\ref{fig:clshr_all_los}). 
In general, $\overline{L}_{cPIL}$ decreases as the detection threshold increases, with about 7~Mm per 25~G averagely. 
For a fixed threshold, the $\overline{L}_{cPIL}$ calculated from higher durations are slightly lower. 
Taking the values obtained at the threshold of 100~G and duration of 3~hours as an example (Figure~\ref{fig:clshr_all_los}(h)), 
the mean value of the $\overline{L}_{cPIL}$ is 42.8~Mm. 
The $84\%$ (16$/$19) of the $\overline{L}_{cPIL}$ are longer than 18.8~Mm (1 $\sigma$ below the average) excepting the two conjoined sPIL/ePIL and a cPIL in AR 12089. 
The two lower cutoff values obtained on both $B_r$ and $B_{los}$ datasets are quite close, we thus take 18~Mm as a reference value when assessing the intensity of the collision. 
Relatively nontrivial correlation still exists between F and $\overline{L}_{cPIL}$ for all thresholds and durations on $B_{los}$. The correlation coefficients range from 0.56 to 0.82 with a confidence level higher than $98\%$. 
When only considering the 9 cases near the disk center, there is also relatively nontrivial correlation. The correlation coefficients are slightly lower than on the full sample at the smaller threshold (50~G to 100~G), and slightly higher at the larger threshold (125~G to 175~G), ranging from 0.51 to 0.91. The confidence levels are lower, ranging from $83.8\%$ to $99.9\%$, which may be because that the sample size of 9 cases is smaller. 
When excluding the two sPIL/ePIL cases, the correlation coefficients also decrease (by around 0.02 to 0.1), but are still nontrivial, 
ranging from 0.54 to 0.77. This also indicates that the two sPIL/ePIL cases do bias the correlation, 
but do not seem to significantly affect the trend indicated by the correlation that intenser activities tend to originate from PILs with longer collisional length. 

Overall, the results obtained at larger thresholds ($\geq$100~G, Figure~\ref{fig:clshr_all_los}(g)-(r)) on $B_{los}$ correlate well with those from $B_r$ dataset. 
For the ones obtained from lower thresholds (50~G and 75~G, Figure~\ref{fig:clshr_all_los}(a)-(f)), the correlation, with coefficients ranging from 0.56 to 0.71, becomes weaker but is still non-ignorable. 
The other difference is that the $\overline{L}_{cPIL}$ of one conjoined sPIL/ePIL 
from AR 11422 becomes significantly larger, e.g., increasing from 10.0~Mm to 48.0~Mm when the threshold changes from 100~G to 50~G at an averaging duration of 3 hours (see also in Section~\ref{subsec:s+epil}). 
The increase of the $\overline{L}_{cPIL}$ again indicates that lower threshold may be more appropriate for the sPIL/ePIL cases when detecting their collision. 
The correlation obtained on $B_{los}$ dataset also supports that there is a trend that intenser activities tend to originate from the PILs with longer collisional length. 

\begin{figure*}
\begin{center}
\epsscale{1.2}
\plotone{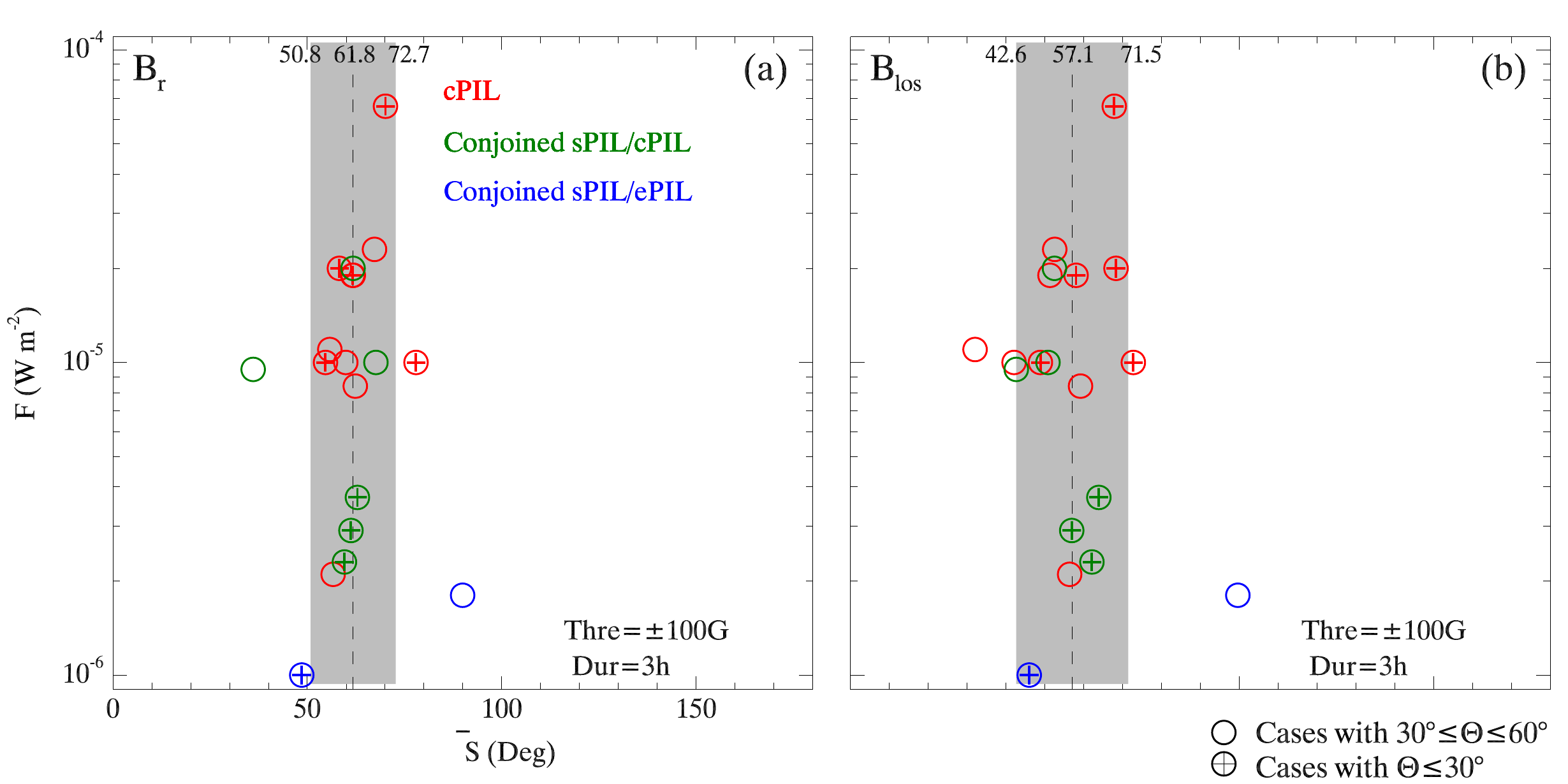}
\caption{
The scatter plots of the mean shear angle ($\overline{S}$) vs. the GOES 1-8~\AA~flux. 
The $\overline{S}$ of each AR is averaged in three hours prior to the first major activity based on the collisional PIL part detected at the threshold of 100~G on $B_r$ dataset in panel (a), and at 100~G on $B_{los}$ dataset in panel (b). 
The vertical lines, shaded regions and labels have similar meaning as the ones in Figure~\ref{fig:clshr_all_br}.   
}\label{fig:flrshr}
\end{center}
\end{figure*}

The correlations between F and the mean shear angle $\overline{S}$ calculated from both datasets at all thresholds and durations show no essential difference. 
We thus show the scatter diagram between $F$ and $\overline{S}$ detected at 100~G and averaged over 3 hours 
in Figure~\ref{fig:flrshr}. 
For the $B_r$ dataset, the average of $\overline{S}$ is around 62$^{\circ}$. 
About 78.9$\%$ of $\overline{S}$ fall into the range of 50.8$^{\circ}$ to 72.7$^{\circ}$ (1$\sigma$ below and over the average). 
Taking the lower cutoff value 
($50^\circ$, around 1 $\sigma$ below the average) as a reference, 
it is found that except one conjoined sPIL/cPIL 
in NOAA AR 11762, the $\overline{S}$ of which is 36.1$^{\circ}$, 
the rest of the PILs all have $\overline{S}$ {very close to or} larger than that.  
For the $B_{los}$ dataset, the mean $\overline{S}$ is around $57^\circ$, close to the one from $B_r$. However, the lower cutoff value (around $43^\circ$) is lower. 
This may be because that $\overline{S}$ for $B_{los}$ is also calculated using the vector field but on the collisional PIL part detected on $B_{los}$ series, which may slightly deviate from the PIL detected on $B_r$.  Except the one $\overline{S}$ of AR 11899, which is $32^\circ$, the rest are all very near to or significantly larger than the lower cutoff value.  
The result suggests that significant shear has been built up at the collisional parts of almost all active PILs prior to the eruptions. 
The flares magnitude exhibits no correlation to $\overline{S}$. 
We also find no clear correlation between the CME velocities and $\overline{L}_{cPIL}$ and $\overline{S}$, so that the results are not shown here.

\section{Discussion and Conclusion} 
\label{sec:discuss}

In this work, 
through tracking 19 ARs from the beginning of their emergence until they produce the first major activities, 
we investigate the formation and properties of the active PILs, i.e., the source PILs of the eruptions. 
We find that none of the active PILs is simply formed by a individual bipole. 
On the contrast, all of them contain non-self PIL parts.  
We further find that the PILs can be classified into three types,  
including 11 cPILs formed due to collision between different bipoles, 
6 conjoined sPIL/cPIL 
which contain the self PIL parts and collisional PIL parts, 
and 2 conjoined sPIL/ePIL 
which are composed of the self PIL parts and external PIL parts. 
Moreover, we find that the flares magnitude is positively correlated with the length of the collisional parts of the active PILs, which holds on both $B_r$ and $B_{los}$ datasets at all thresholds and averaging durations we investigated.

For the 11 cPILs, collision between nonconjugated opposite-sign polarities develops at all of the PILs prior to the first major activities of the ARs. 
Observations reveal that the collision accompanied by shearing and flux cancellation is driven by the self-separation of different bipoles when they emerge, 
consistent well with the ``collisional shearing'' scenario proposed by \citet{Chintzoglou_2019}.  
The length of the collisional parts clearly show a trend of increase from the emergence start to the activity onset, 
while the non-potential shear angles also show a trend of increase although not that dramatic, further supporting the ongoing collisional shearing. Taking the collisional lengths detected on $B_r$ series at 100~G and averaged in 3 hours (prior to the flare) as the example, 10 of them are longer than the reference value of significant collision (18~Mm).

For the 6 conjoined sPIL/cPIL, 
collision also develops and grows at part of the PILs prior to the activities, 
with shearing and flux cancellation observed. 
All of the averaged collisional lengths are longer than 18~Mm. 
In the ARs containing this type of PILs, 
collision may not only occur between nonconjugated opposite-sign polarities, 
but also occur between same-sign polarities of different bipoles. 
This kind of collision may further shear the field lines through moving the footpoints of the field lines. For example, in the sPIL/cPIL case of AR 11776, the negative polarity NA pushes another negative polarity NB to the west, which makes NB deviate from its original northward direction, and apparently can shear the field lines connecting NB and PB. This indeed also is a kind of photospheric motions (see also in Section~\ref{sec:intro}) driven by the collision. 
  
For the 2 conjoined sPIL/ePIL, 
collision signatures are also observed. The collisional length is short at the threshold of 100~G.  
Considering that the magnetic field of the two ARs is not as strong as others, it might be more appropriate to use lower thresholds (e.g., 50~G) on low noise $B_{los}$ when detecting their collision. 
When lowering the threshold to 50~G, 
the collision length of AR 11422 becomes significantly longer until the flare. 
That of AR 11870 evolves dramatically, being longer at the early emergence phase and shorter near to the flare. 
As there are only two sPIL/ePIL cases, 
we cannot draw  general conclusion on the collision feature of this type of PILs. 
We have strictly followed the criteria (see Section~\ref{sec:obs_ana}) when selecting the cases from the list of 423 emerging ARs in~\citet{Kutsenko_2019}. 
As also mentioned in~\citet{Chintzoglou_2019}, the criteria of ``emerging and producing major eruptions on the visible disk'' are strict 
thus have screened out most of the cases. 
The few sPIL/ePIL cases here is more likely to be an indicator of how often such cases occur. 

Overall, at all active PILs we studied, collisional signatures are found developing prior to the first major activities of the ARs, 
with 16 of them having averaged collisional length longer than 18~Mm.  
Here we define 18~Mm (detected at 100~G) 
as a reference value when assessing the significance of the collision, while that in~\citet{Chintzoglou_2019} is defined as 40 Mm. 
Nevertheless, the latter is obtained based on two large, productive ARs, whilst 18~Mm 
here is calculated as 1 $\sigma$ below the averaged collisional length of all active PILs, 
which are got from 19 ARs owning various sizes. 
Thus the two values may not be conflicting, and the value of 18~Mm 
may be more general.

\begin{figure*}
\begin{center}
\epsscale{1.25}
\plotone{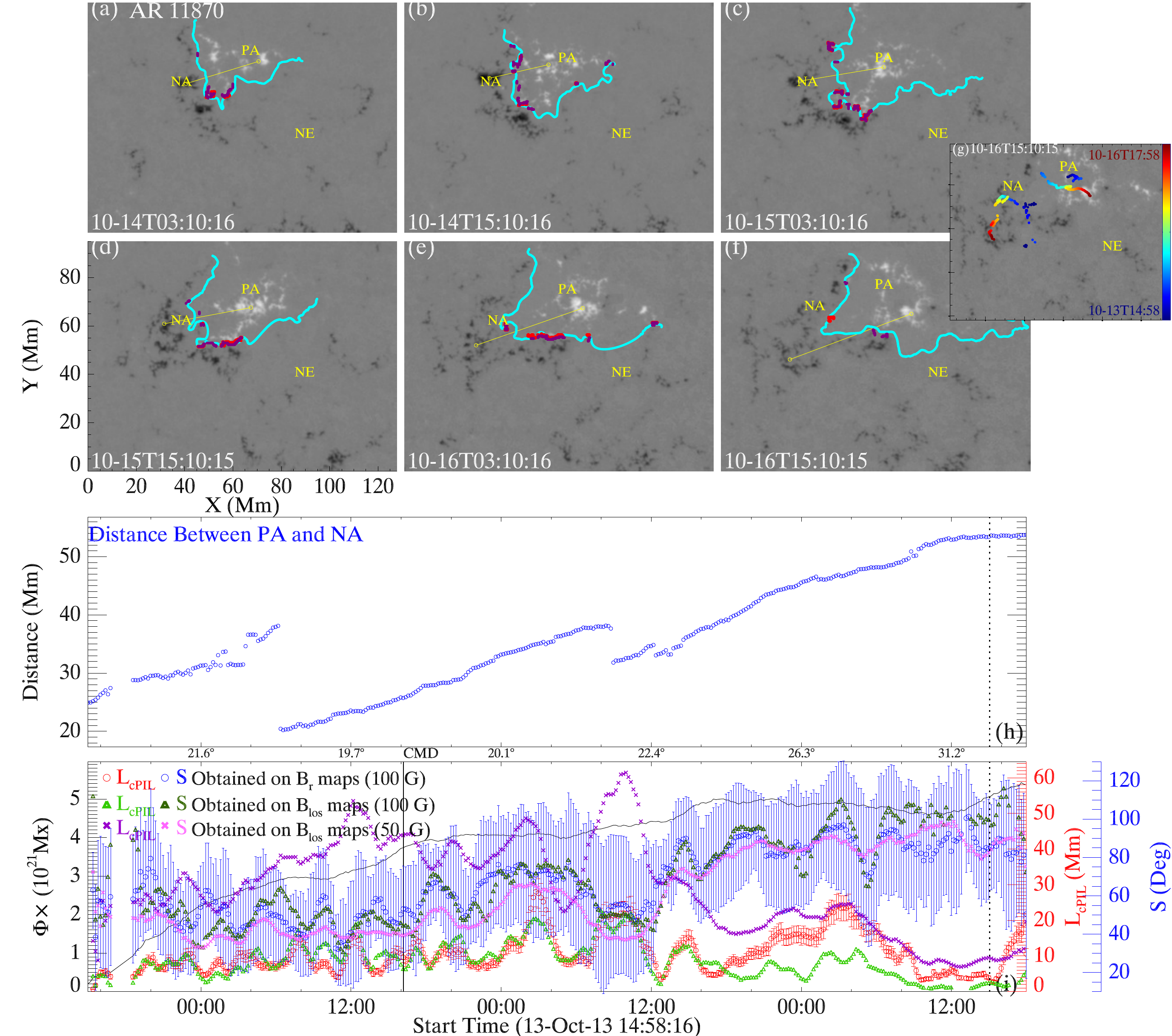}
\caption{
The evolution of NOAA AR 11870. Same layout as Figure~\ref{fig:11422_b}. 
An animation of the magnetograms lasting from 2013-10-13T06:10 to 2013-10-16T17:58 is available online. 
}\label{fig:11870_b}
\end{center}
\end{figure*}

Note that 18~Mm is just a reference from a statistical perspective. 
For the three cases owning $\overline{L}_{cPIL}$ shorter than 18~Mm, 
the small collision at the cPIL one in AR 12089 is more likely resulted from the small size of the AR (around $1.2\times 10^{21}$~Mx) since the first major activity occurs only 4 hours after the emergence onset. 
The collision is relatively significant for the small AR, playing an important role in driving the eruption. 
For the conjoined sPIL/ePIL one in AR 11422, the collision before the flare, 
which also mainly occurs between the nonconjugated polarities, 
increases to as long as 48~Mm when lower the detection threshold to 50~G, 
indicating that the collision is rather significant. 
For the other one, i.e., the conjoined sPIL/ePIL in AR 11870, 
the collisional length before the flare is short (below 10~Mm) even at the lower threshold, 
but is rather significant (with $L_{cPIL}$ as high as 60~Mm) in the early emergence phase. 
Different from the other ARs, quite a part of the early collision in AR 11870 occurs between the conjugated polarities (e.g., Figure~\ref{fig:11870_b}(b)), which may be resulted from the small moving dipoles (the conjugated polarities of which are very close) during the emergence.  
The early collision may have transferred the emerging field higher into the corona, 
which may interact with other higher, preexisting magnetic field,  playing a role in producing the eruptions.  
A similar scenario is discussed in~\citet[][and references therein]{Schmieder_2014}. 

The positive correlation between the flares magnitude and the length of the collisional parts of the PILs  suggests that the intenser activities tend to originate from the longer collisional PILs, further suggesting that the collision plays the important role in generating the large activities. Considering that the field at almost all collisional PILs is significantly sheared before the eruptions, the longer collisional PILs indicate that more sheared magnetic flux is involved into the cancellation, thus a larger amount of magnetic free energy may be available to consume during the eruptions.

As the flares and CMEs are suggested to be closely related to the eruptions of the magnetic flux ropes~\citep[e.g.,][]{Shibata_1995}, 
their source locations may give hint to the origin of the coronal flux ropes. 
For the two generally accepted scenarios of coronal flux rope formation~\citep[][and reference therein]{Chengx_2017}, 
the bodily emergence from the solar interior should produce a bipolar region on the photosphere~\citep[e.g.,][]{Fan_2004}, while formation 
in the solar atmosphere usually doesn't support simple bipole configuration.    
In the two productive ARs in~\citet{Chintzoglou_2019}, 
none of the eruptions originated from the self PIL of a single bipole. The authors thus suggested that no flux rope formed or emerged above the sPIL, and the bodily emergence of the flux rope may be rare. 
In our sample, 
lack of the activities originated from the single sPIL also supports this point. 
It is also consistent with other reported observations, in which the filament (a proxy of the flux rope) formed above the self PIL is found to be rare~\citep{MacKay_2008}.

To summarize,  
in all of the emerging ARs we studied, the collision develops at the active PILs prior to the first major activities from the ARs, and in at least $84\%$ (16 of 19) 
of them, the collisional shearing is quite significant.   
Furthermore, the magnitude of the flares is positively correlated with the length of the collisional parts of the PILs.  
From the statistical perspective, 
the results consolidate that the bipole-bipole interactions during the flux emergence play the important role in driving the major solar activities. Moreover, the length of the collisional PIL may be a promising indicator in forecasting the major solar activities.

\begin{longrotatetable}
\begin{longtable}{c|c|cc|ccc|c|c|ccc}

\caption{Information of the ARs, their first major activities and the active PILs}\label{tb:props}
\\
\hline
\hline
No. & \multicolumn{3}{c}{ARs} & \multicolumn{4}{c}{First major activities} & \multicolumn{4}{c}{Active PIL} \\
\hline
    &  NOAA  & \multicolumn{2}{c|}{Flux emergence onset$^{\hyperref[a1]{a}}$} & \multicolumn{3}{c|}{Flare$^{\hyperref[b0]{b}}$} & CME & Collisional & PIL & $\overline{L}_{cPIL}$ $^{\hyperref[e1]{f}}$ &  $\overline{S}^{\hyperref[f1]{g}}$  \\ 
    & No. & Time & Location & Start & Location & Class & Speed $^{\hyperref[b1]{c}}$ &  shearing  & type $^{\hyperref[d1]{e}}$ & & \\ 
    & &   &   &  &  &  & (km s$^{-1}$) & case $^{\hyperref[c1]{d}}$ &  & &  \\
\hline
1   &  11081    & 2010-06-11T02:24 & N24W32 (43$^{\circ}$)  & 2010-06-12T00:30 & N23W43 (52$^{\circ}$) & M2.0   & 486  & B & S/C  & 49.5$\pm$12.8  & 61.7$\pm$6.9 \\
2   &  11158    & 2011-02-10T17:48 & S20E47 (48$^{\circ}$)  & 2011-02-13T17:28 & S19W03 (15$^{\circ}$) & M6.6   & 373  & A & C    & 62.7$\pm$3.8  & 70.1$\pm$1.1 \\
3   &  11162    & 2011-02-17T14:00 & N18E15 (29$^{\circ}$)  & 2011-02-18T10:23 & N18E02 (26$^{\circ}$) & M1.0   &      & B & C    & 87.5$\pm$6.4  & 77.9$\pm$4.2 \\
4   &  11422    & 2012-02-18T10:00 & N16E25 (33$^{\circ}$)  & 2012-02-19T08:41 & N17E10 (26$^{\circ}$) & C1.0   & 238  & B & S/E  & 11.7$\pm$2.2   & 48.6$\pm$7.2 \\
5   &  11440    & 2012-03-20T04:00 & S26W00 (18$^{\circ}$)  & 2012-03-21T12:38 & S27W20 (25$^{\circ}$) & C2.9   & 387  & A & S/C  & 24.9$\pm$6.5  & 61.2$\pm$4.3 \\
6   &  11466    & 2012-04-20T21:24 & N13E58 (61$^{\circ}$)  & 2012-04-27T08:15 & N12W30 (35$^{\circ}$) & M1.0   & 365  & B & S/C  & 22.3$\pm$6.4  & 67.6$\pm$5.8 \\
7   &  11620    & 2012-11-24T23:00 & S13W02 (15$^{\circ}$)  & 2012-11-27T21:05 & S14W41 (43$^{\circ}$) & M1.0   &      & B & C    & 60.1$\pm$4.5  & 59.9$\pm$1.7 \\
8   &  11675    & 2013-02-15T22:12 & N12E47 (51$^{\circ}$) & 2013-02-17T15:45 & N12E22 (30$^{\circ}$) & M1.9   &      & A & C    & 39.4$\pm$7.4  & 61.7$\pm$3.4 \\
9   &  11762    & 2013-06-01T03:36 & S29E04 (29$^{\circ}$) & 2013-06-03T07:03 & S27W21 (38$^{\circ}$) & C9.5   & 429$^{\hyperref[dag1]{\dag}}$  & B & S/C & 61.0$\pm$7.3  & 36.1$\pm$2.1 \\
10  &  11776    & 2013-06-18T07:12 & N11E15 (18$^{\circ}$)  & 2013-06-19T00:50 & N10E03 (10$^{\circ}$) & C2.3   & 287  & A & S/C  & 20.0$\pm$1.6  & 59.5$\pm$5.9 \\
11  &  11817    & 2013-08-10T09:36 & S21E44 (51$^{\circ}$) & 2013-08-11T21:47 & S20E25 (36$^{\circ}$) & C8.4   & 110  & A & C    & 69.1$\pm$4.6 & 62.3$\pm$2.0 \\
12  &  11870    & 2013-10-13T05:00 & S14E19 (28$^{\circ}$) & 2013-10-16T15:03 & S15W29 (33$^{\circ}$) & C1.8   & 514  & B & S/E  & 4.3$\pm$1.6   & 89.9$\pm$15.9\\
13  &  11891    & 2013-11-05T12:48 & S18E09 (25$^{\circ}$) & 2013-11-08T09:22 & S17W28 (37$^{\circ}$) & M2.3   & 207  & A & C    & 41.8$\pm$3.8  & 67.3$\pm$2.4 \\
14  &  11899    & 2013-11-15T15:24 & N10E32 (47$^{\circ}$) & 2013-11-23T02:20 & N14W56 (56$^{\circ}$) & M1.1   & 406  & A & C    & 61.7$\pm$6.5  & 55.8$\pm$3.4 \\
15  &  11928    & 2013-12-16T08:24 & S16E31 (35$^{\circ}$) & 2013-12-22T08:05 & S19W51 (51$^{\circ}$) & M1.9   & 231  & A & C    & 70.9$\pm$5.7  & 61.8$\pm$1.6 \\
16  &  11946    & 2014-01-04T05:12 & N09E49 (51$^{\circ}$) & 2014-01-07T03:49 & N07E08 (13$^{\circ}$) & M1.0   &      & B & C    & 19.5$\pm$3.4  & 54.7$\pm$2.4 \\
17  &  12017    & 2014-03-23T22:36 & N03E53 (54$^{\circ}$) & 2014-03-28T19:04 & N11W21 (25$^{\circ}$) & M2.0   & 420  & B & C    & 44.3$\pm$2.1  & 58.2$\pm$2.9 \\
18  &  12085    & 2014-06-05T22:36 & S20E39 (44$^{\circ}$) & 2014-06-09T01:14 & S20E00 (21$^{\circ}$) & C3.7   & 417  & B & S/C  & 30.3$\pm$5.8  & 62.9$\pm$3.9 \\
19  &  12089    & 2014-06-10T20:00 & N18E32 (35$^{\circ}$) & 2014-06-10T23:46 & N17E29 (34$^{\circ}$) & C2.1   & 343  & A & C    & 12.9$\pm$7.4  & 56.6$\pm$10.6 \\
\hline 

\end{longtable}


$^a$ The onset time and location 
of the flux emergence of the ARs.  The locactions are the Stonyhurst coordinates (outside the brackets) and the disk-cented angles (in the brackets). The time and Stonyhurst locations 
are cited from~\citet{Kutsenko_2019}.\label{a1}\\ 
$^b$ The onset time, location (Stonyhurst and disk-centered angles) and class of the flares. Referred to the GOES flare catalog\textsuperscript{\ref{ft:f1}}. \label{b0} \\
$^c$ CME velocities provided by the {\it SOHO}/LASCO CME catalog. The blank in this column means no CME is associated with the flare.\label{b1} \\
$^{\dag}$ CME velocity provided by the {\it STEREO}/SECCHI COR2 CME catalog.\label{dag1} \\
$^d$:Collisional shearing 
case of the two colliding bipoles. 
Case A and case B stand for simultaneous and sequential collisional shearing, 
respective. \label{c1} \\
$^e$ Type of the PILs. Here we use ``C'', ``S/C'' and ``S/E'' to represent the cPIL, conjoined sPIL/cPIL 
and conjoined sPIL/ePIL 
for convienience.  \label{d1} \\
$^f$ $\overline{L}_{cpil}$: Length of the collisional parts of the PILs averaged in three hours prior to the activities. The standard deviations are taken as the errors. The collisional parts of the PILs are detected at the threshold of 100~G on $B_r$ dataset. \label{e1} \\
$^g$ $\overline{S}$: Shear angles of the field at the collisional parts of the PILs, which are averaged in three hours prior to the activities. The standard deviations are taken as the errors. \label{f1} \\

\end{longrotatetable}

\clearpage

\acknowledgments{
We acknowledge the use of data from {\it SDO}, {\it{GOES}}, and {\it SOHO}. 
We thank our anonymous referee for his/her constructive report which helps to improve the paper significantly. 
We thank Kutsenko A., Abramenko V., and Pevtsov A. for providing the list of the emerging ARs. 
L.L. is supported by NSFC (11803096), the Fundamental Research Funds for the Central Universities (19lgpy27) and the Open Project of CAS Key Laboratory of Geospace Environment.
Y.W. is supported by NSFC (41574165, 41774178). 
Z.Z. is supported by the Open Project of CAS Key Laboratory of Geospace Environment. 
J.C. is supported  by NSFC (41525015, 41774186). 
}

\clearpage
\bibliography{Emerging_AR}
\bibliographystyle{aasjournal}

\end{document}